\documentclass[letterpaper,fleqn,usenatbib]{mn2e}

\pdfoutput=1
\usepackage{times}
\long\def\symbolfootnote[#1]#2{\begingroup%
\def\thefootnote{\fnsymbol{footnote}}\footnote[#1]{#2}\endgroup}
\usepackage[T1]{fontenc}
\usepackage{ae,aecompl}

\usepackage{graphicx}	
\usepackage{amsmath}	
\usepackage{amssymb}	
\usepackage{multirow}
\usepackage{natbib}
\usepackage{subfigure}
\usepackage{xcolor}
\usepackage{ulem}

\defcitealias{read16}{R16b}

\title[The upper bound on the lowest mass halo]{The upper bound on the lowest mass halo}

\author[P. Jethwa et al.]{
P. Jethwa,$^{1}$\thanks{E-mail: pj253@ast.cam.ac.uk}
D. Erkal$^{1}$ \& V. Belokurov$^{1}$
\\
$^{1}$Institute of Astronomy, University of Cambridge, Madingley Road, Cambridge CB3 0HA, UK\\
}

\date{Accepted 2017 September 1. Received 2017 August 10; in original form 2016 December 21}

\pubyear{2017}

\begin{document}
\label{firstpage}
\pagerange{\pageref{firstpage}--\pageref{lastpage}}
\maketitle

\begin{abstract}
We explore the connection between galaxies and dark matter halos in the Milky Way (MW) and quantify the implications on properties of the dark matter particle and the phenomenology of low-mass galaxy formation.
This is done through a probabilistic comparison of the luminosity function of MW dwarf satellite galaxies to models based on two suites of zoom-in simulations.
One suite is dark-matter-only while the other includes a disk component, therefore we can quantify the effect of the MW's baryonic disk on our results.
We apply numerous Stellar-Mass-Halo-Mass (SMHM) relations allowing for multiple complexities: scatter, a characteristic break scale, and subhalos which host no galaxy.
In contrast to previous works we push the model/data comparison to the faintest dwarfs by modeling observational incompleteness, allowing us to draw three new conclusions.
Firstly, we constrain the SMHM relation for $10^2<M_*/M_\odot<10^8$ galaxies, allowing us to bound the peak halo mass of the faintest MW satellite to $M_\mathrm{vir}>2.4\times10^8M_\odot$ ($1\sigma$).
Secondly, by translating to a Warm Dark Matter (WDM) cosmology, we bound the thermal relic mass $m_\mathrm{WDM}>2.9$ keV at 95\% confidence, on a par with recent constraints from the Lyman-$\alpha$ forest.
Lastly, we find that the observed number of ultra-faint MW dwarfs is in tension with the theoretical prediction that reionisation prevents galaxy formation in almost all $10^8M_\odot$ halos.
This can be tested with the next generation of deep imaging surveys.
To this end, we predict the likely number of detectable satellite galaxies in the Subaru/HSC survey and the LSST.
Confronting these predictions with future observations will be amongst our strongest tests of WDM and the effect reionisation on low-mass systems.
\end{abstract}



\section{Introduction}
\label{sec:intro}

The total matter content in the Universe is known with great precision \citep[see e.g.][]{Planck2016}, and must be apportioned to its individual parts, thus providing a powerful -- albeit indirect -- means of weighing objects whose masses are not accessible otherwise.
Applied to galaxies, this concept manifests as the technique known as abundance matching \citep{vale04}.
In this, we infer a galaxy's virial mass subject to its dark matter (DM) halo being part of a hierarchy described by a mass function (MF), and postulating monotonicity between halo mass and galaxy stellar mass.
This is useful because virial masses can rarely be measured directly, since baryons are typically confined to the very bottom of the galaxy's potential well.
This observational hurdle to inferring the virial mass affect the smallest galaxies, the dwarfs, the most: their DM halos are puny and lose gas quickly, thus leaving only the very central parts lit up with stars.
Moreover, the tiniest of the dwarfs detected to date are all satellites of much more massive hosts, and hence have all been on the receiving end of tidal shredding.
Nevertheless, by re-framing the galaxy-halo connection in terms of the peak virial mass attained over the course of a halo's evolution \citep{conroy06}, one can probe the behaviour of the DM subhalo MF in the low-mass regime and to test the physics of star-formation in the early Universe.

In the Milky Way (MW), satellite objects with extremely low luminosities have recently been discovered \citep[e.g.][]{UMa, W1}.
Using spectroscopic follow-up observations, many of these have been shown to contain substantial amounts of DM \citep[see e.g.][]{simongeha}. 
The Galactic dwarf census therefore offers an opportunity to scrutinise the relationship between the DM halos and the dwarf galaxies across a wide range of dark and baryonic masses.
Such a study is however only possible if the number of satellites observed is corrected for the detection efficiency.
As comprehensively demonstrated in \citet{koposov08}, the correction required is a steep function of the dwarf's luminosity and distance.
This strikes a twofold blow to the quality of the local dwarf census: the uncertainty due to the loosely constrained frequency of low-luminosity satellites is amplified by the poorly known distance profile of the DM subhalos.
The problem is exacerbated by the fact that little effort has gone so far into establishing a rigorous statistical framework for the comparison between luminosity function (LF) observations and model predictions.

In a somewhat naive manner, one could attempt such a comparison by simply counting the number of dwarf satellites around the Galaxy and the number of subhalos in a Cold Dark Matter (CDM) numerical simulation.
This admittedly oversimplified experiment would register a striking mismatch between the two numbers, a phenomenon currently known as the ``missing satellite'' problem \citep{moore99,klypin99}.
Multiple solutions to this problem have been offered, including scenarios where the DM MF is truncated below a certain mass.
This is an attractive proposition, since the mass threshold can be linked directly to the mass of the particle sourcing the DM.
Representing relatively light, and therefore, fast moving particles, so-called Warm Dark Matter (WDM) has recently been hotly debated in the literature \citep[see e.g.][]{boyarsky09}.
An example of the application of the WDM idea to the sub-Galactic scales can be found in \citet{lovell14} who modify the initial conditions of their DM-only simulations to represent WDM particles with masses between 1.5 and 2.3 keV.
In this experiment, the $z=0$ analogs of the MW start to show signs of depletion in the subhalo MF at $7\times10^9 M_{\odot}$ and $2\times10^9 M_{\odot}$ correspondingly.
When comparing their prediction to the observations, \citet{lovell14} did not apply any correction for satellite detection efficiency, rendering their constraints somewhat incomplete.

In CDM, on the other hand, the subhalo MF has been shown not to deviate much from a power-law down to masses resolvable in the state-of-the-art collisionless simulations, i.e. to values as low as
$10^{-6} M_{\odot}$ \citep[see e.g.][]{diemand2005}.
No one, of course, expects many of these $10^{15}$ CDM subhalos predicted to surround the MW today to have formed any appreciable number of stars.
Below a certain mass, subhalos fail to accrete, cool and hold on to enough gas to produce a detectable stellar counterpart \citep[e.g.][]{rees86, dekel1986, efstathiou92, bullock2000}.
Therefore, even though the CDM subhalo MF may run uninhibited across almost two decades in mass, a cut-off -- or a turn-around -- in the dwarf galaxy MF is predicted to exist.
However, the lowest mass of a subhalo to host a dwarf has not been pinned down yet.
For example, most recently, \citet{shen2014,wheeler2015,sawala16} all used hydrodynamic cosmological zoom-in simulations to claim that below a subhalo virial mass of $10^9 M_{\odot}$, dwarf galaxy
formation is strongly suppressed due to i) the emergence of the ionising background around the dwarf and ii) the stellar feedback within the dwarf.
This can be contrasted with the study by \citet{blandhawthorn15}, where a much higher effective resolution is employed to include the appropriate gas physics in a non-cosmological case study of the birth of a dwarf galaxy.
As a result, it is demonstrated that subhalos with masses on the order of $10^7 M_{\odot}$ can host ultra-faint dwarfs galaxies.

In this paper, we constrain the mass of the smallest DM subhalo to host a dwarf galaxy via a technique that has features in common with both semi-analytic galaxy formation models and abundance matching.
The motivation behind our analysis is not only to compare the Galactic satellite data against the predictions from the variety of new cutting edge numerical simulations mentioned above: we strive to improve on the method itself.
Our {\it subhalo-dwarf mapping} links dwarf galaxies with their DM hosts, without necessarily trying to explicitly reproduce the underlying physics, or preserve mass/luminosity ranking.
The analysis presented here differs from those reported in the literature in a number of ways.
For example, compared to the extensive study of the MW satellite LF presented in \citet{koposov09}, instead of the merger-tree based subhalo families, we use those produced in cosmological zoom-in simulations of MW-like hosts.
In this regard, our approach is similar to that of \citet{garrison-kimmel_elvis}, however here we do not limit our inference to satellites with stellar masses above $10^5M_{\odot}$, but model the entirety of the Galactic satellite population.
To capture the physics of star formation in low mass halos, we do not resort to semi-analytic modeling as presented in e.g. \cite{maccio10} or \citet{font11} but instead test a range of Stellar-Mass-Halo-Mass (SMHM) relations with varying degree of sophistication.

Furthermore, none of the works above include the action of the MW's baryonic disk.
This is unfortunate as the disk has been shown to play an important role in sculpting the properties of the Galactic satellites \citep[e.g.][]{penarrubia02,penarrubia10,donghia_et_al_2010,errani_etal_2017}.
In simplest terms, the disk depletes the dwarf satellite population by a factor of several via enhanced tidal shredding.
Note that the exact processes at work are probably much more complex and involve the possibility of the DM halo contraction in response to the disk, which would in turn increase the satellite disruption further.
Additionally, as several numerical experiments have now demonstrated, a significant fraction of model satellites can be wiped out due to the coupling between the stellar feedback, which would loosen the dwarf DM distribution, and the repeated battering by the disk shocks \citep[see e.g.][]{geen13,brooks14,wetzel16}.
Here, for the first time, we include the effects of the baryonic disk as part of the subhalo-dwarf mapping.
Our simulations are DM-only, but are run twice, once to produce a fiducial set of subhalos and once with an analytic disk grown within the host halo.
As we show below, the effect of the disk is not only to reduce the normalisation of the subhalo radial density profile but also to change its shape.
Thus, the inclusion of the disk is crucial, given that the predicted number of the Galactic dwarf satellites depends sensitively on the assumed subhalo radial profile \citep[see][]{koposov08,tollerud08,belokurov13,hargis14}.

As explained earlier, the observed MW dwarf satellite LF can be used to place constraints on the WDM particle mass \citep[see e.g.][]{maccio10dm,polisensky11,kennedy14}.
As evidenced in the current literature, this inference is normally carried out within the framework of the chosen semi-analytic model of dwarf galaxy formation.
Additional systematic uncertainties include the treatment of (or neglect of) the selection effects due to dwarf satellite detectability as well as the poorly known mass of the MW galaxy.
In the analysis presented here, we provide new limits on the WDM particle mass using a more streamlined and self-consistent approach.
Instead of a semi-analytic galaxy formation model, we find the best-performing subhalo-dwarf mapping recipe, which itself is constrained using the observed dwarf satellite LF (having taken into account the dwarf detection efficiency).
As mentioned above, there is a further improvement due to the inclusion of the effects of the baryonic disk.
With regards to the MW mass, we probe a wide range of host masses allowed by observations.

It is our aim not only to gauge the total DM peak mass of the faintest dwarf satellite within the CDM paradigm, but also to give a robust estimate of its associated confidence interval.
Achieving this goal is impossible without the introduction of some sort of goodness-of-fit measure, and is truly tortuous outside of the probabilistic framework.
Nevertheless, the majority of the works discussed above choose not to quantify the performance of their LF models.
The only exception, to our knowledge, is the analysis of \citet{garrison17} who introduce goodness-of-fit measure similar in spirit to the Kolmogorov-Smirnov statistic used to compare two distributions. While this quantity helps to rank different models according to their performance, its utility for the subsequent uncertainty evaluation is unclear.
Here, for the first time, we provide a fully probabilistic treatment of the problem.

This Paper is organized as follows. Section~\ref{sec:method} gives the details of the subhalo-dwarf mapping models as well as the statistical method employed to carry out the inference.
Section~\ref{sec:results} presents the results of the LF fitting with the mapping recipes described.
It also gives predictions for the number of discoverable satellites of the MW in upcoming surveys (LSST and Subaru/HSC).
Implications for SMHM relation and constraints on the lowest mass halo are presented in Section~\ref{sec:SMHM}.
In Section~\ref{sec:disk_nodisk} we discuss the effect of the MW disk on our results, then WDM constraints are discussed in Section~\ref{sec:wdm}.
Finally we discuss caveats and implications in Section~\ref{sec:discuss}, then conclude in Section~\ref{sec:conclude}.

\section{Method}
\label{sec:method}

The outline of our subhalo-dwarf mapping method, and the structure of this section, is as follows: we perform cosmological $N$-body zoom-in simulations (Section~\ref{sec:sims}), then assign stellar mass to dark matter subhalos using parameterised SMHM relations (Sections~\ref{ssec:AM_proxy} and~\ref{ssec:mstarmods}).
We then model observational biases, and perform a Bayesian comparison with the observed distribution of MW satellites (Sections~\ref{ssec:data} and~\ref{ssec:probmod}), and finally verify and optimise our method using fake data tests (Sections~\ref{ssec:fakedata}).

\subsection{Simulations}
\label{sec:sims}

In order to study the statistical properties of subhalos around a MW-like galaxy, we run a suite of high resolution cosmological $N$-body zoom-in simulations. These simulations are run with the $N$-body part of \textsc{gadget-3} which is similar to \textsc{gadget-2} \citep{springel_2005}. As we descrbe below, in addition to running collisionless $N$-body simulations, we also run the same zoom-in simulations where we follow the MW-like halo and insert a disk potential to account for the disk's effect on the subhalos as found in \cite{donghia_et_al_2010}. These disks significantly deplete the subhalos within 30 kpc and 100 kpc by a factor of 2-4 and $\sim$2 respectively. For the cosmological parameters we use the results from Planck 2013 \citep{planck_2013} with $h=0.679$, $\Omega_b = 0.0481$, $\Omega_0 = 0.306$, $\Omega_\Lambda = 0.694$, $\sigma_8 = 0.827$, and $n_s = 0.962$. 

The zoom-in strategy broadly follows \cite{onorbe_etal_2014} and initial conditions are generated with \textsc{music} \citep{hahn_abel_2011}. First, a low resolution simulation is run in a 50 $h^{-1}$ Mpc box with 512$^3$ particles from $z=50$ to $z=0$ in order to find isolated MW analogues. The isolated MW analogues are chosen with similar criteria to those in \cite{garrison-kimmel_elvis}, with virial masses between $7.5\times 10^{11} M_\odot - 2 \times 10^{12} M_\odot$, no major merger since $z=1$, and no haloes with mass greater than half the MW analogue's mass within $2{h}^{-1}$\,Mpc. For each MW analogue, we run a medium resolution zoom-in simulation whose highest resolution region has an effective resolution of 4096$^3$ and a particle mass of $1.54\times 10^5 h^{-1} M_\odot$. We use \textsc{music} to generate the initial conditions for the zoom-in using the ellipsoidal bounding of all particles within 7.5 virial radii of the MW analogue at $z=0$ in the low resolution simulation. For one of the analogues, we also run a high resolution zoom-in with an effective resolution of 8192$^3$ and a particle mass of $1.92\times 10^4 h^{-1} M_\odot$. For the initial conditions in the high resolution setup, we identified all particles within 7.5 virial radii of the MW analogue in the medium resolution simulation and once again use the ellipsoidal bounding region with \textsc{music}. All of the zoom-ins are uncontaminated by low resolution particles within 1 $h^{-1}$ Mpc of the MW analogue.

The softening lengths were selected using the criteria in \cite{power_etal_2003}. For the medium and high resolution zoom-ins, this gave a softening of 360 pc and 144.8 pc respectively for the highest resolution particles. The lower resolution particles were spread over three particle types with successively larger softenings. For the zoom-in simulations, these softenings were comoving until $z=3$, after which they became physical. However, for the low resolution simulation, the softenings were kept comoving. For the zoom-in simulations, snapshots were output approximately every 150 Myr with 104 snapshots saved for each simulation. Halo finding was done with \textsc{rockstar} \citep{rockstar} and the merger trees were constructed with \textsc{consistent trees} \citep{consistent_trees}.
We discuss the effect of using different halo finders in Section~\ref{ssec:robust_test}.
Finally, for ease in analysis, the halo catalogs, trees, and snapshots are stored in an SQL database. In Table~\ref{tab:halo_properties} we give the properties of the halos in our suite. 

\begin{table}
\centering
\caption{ Properties of halos at $z=0$.   }
\begin{tabular}{|c|c|c|c|c|}
Simulation & Disk? & M$_{\rm vir}$ ($ 10^{10} M_\odot$) & $c_\mathrm{vir}$ & $r_s$ (kpc) \\ \hline \hline
\multirow{2}{*}{1}	&	N	&	136	&	8.9	&	32.9	\\
	&	Y	&	139	&	13.8	&	21.4	\\ \hline 
\multirow{2}{*}{2}	&	N	&	89	&	12.2	&	20.9	\\
	&	Y	&	92	&	12.7	&	20.2	\\ \hline 
\multirow{2}{*}{3}	&	N	&	89	&	12.8	&	19.8	\\
	&	Y	&	90	&	19.0	&	13.4	\\ \hline 
\multirow{2}{*}{4}	&	N	&	111	&	7.6	&	35.8	\\
	&	Y	&	110	&	8.9	&	30.4	\\ \hline 
\multirow{2}{*}{5}	&	N	&	91	&	9.8	&	26.1	\\
	&	Y	&	91	&	13.6	&	18.8	\\ \hline 
\multirow{2}{*}{6}	&	N	&	157	&	13.7	&	22.4	\\
	&	Y	&	163	&	16.5	&	18.8	\\ \hline 
\multirow{2}{*}{7}	&	N	&	107	&	6.1	&	44.0	\\
	&	Y	&	100	&	12.7	&	20.8	\\
\end{tabular}
\label{tab:halo_properties}
\end{table}

Once the initial zoom-in is complete, the MW progenitor is identified at $z=3$. We have modified \textsc{gadget-3} so that we can track the main halo on the fly. It is tracked by identifying the center of mass position and velocity within 200 kpc of the potential minimum every $\Delta a = 0.01$. At each timestep between computing the potential minimum, the expected position of the halo is computed and the center of mass position and velocity within 2.5 kpc of the expected position are re-evaluated. This method accurately tracks the halo and the difference in position compared to that reported by the halo finder is less than a softening length. During the re-simulation, we include the force from a Miyamoto-Nagai disk \citep{miyamoto-nagai}. The disk is adiabatically grown from $z=3$ to $z=1$ with mass increasingly approximately linearly in time, $M(a)\propto (a-a_0)^\frac{3}{2}$, where $a_0$ is the scale factor at which the disk is initialised, i.e. $a_0 = 1/4$. After $z=1$, the disk mass and size are kept fixed with $M=8 \times 10^{10} M_\odot$, $a=3$ kpc, and $b=300$ pc. The disk orientation is kept fixed with a normal in the $z$ direction; this direction is effectively random since the galaxy's angular momentum can point in any direction. As a final check, we also compared the trajectory of the halo with the disk with the fiducial halo and found that their tracks are almost identical. 

\begin{figure}
\includegraphics[width=\columnwidth]{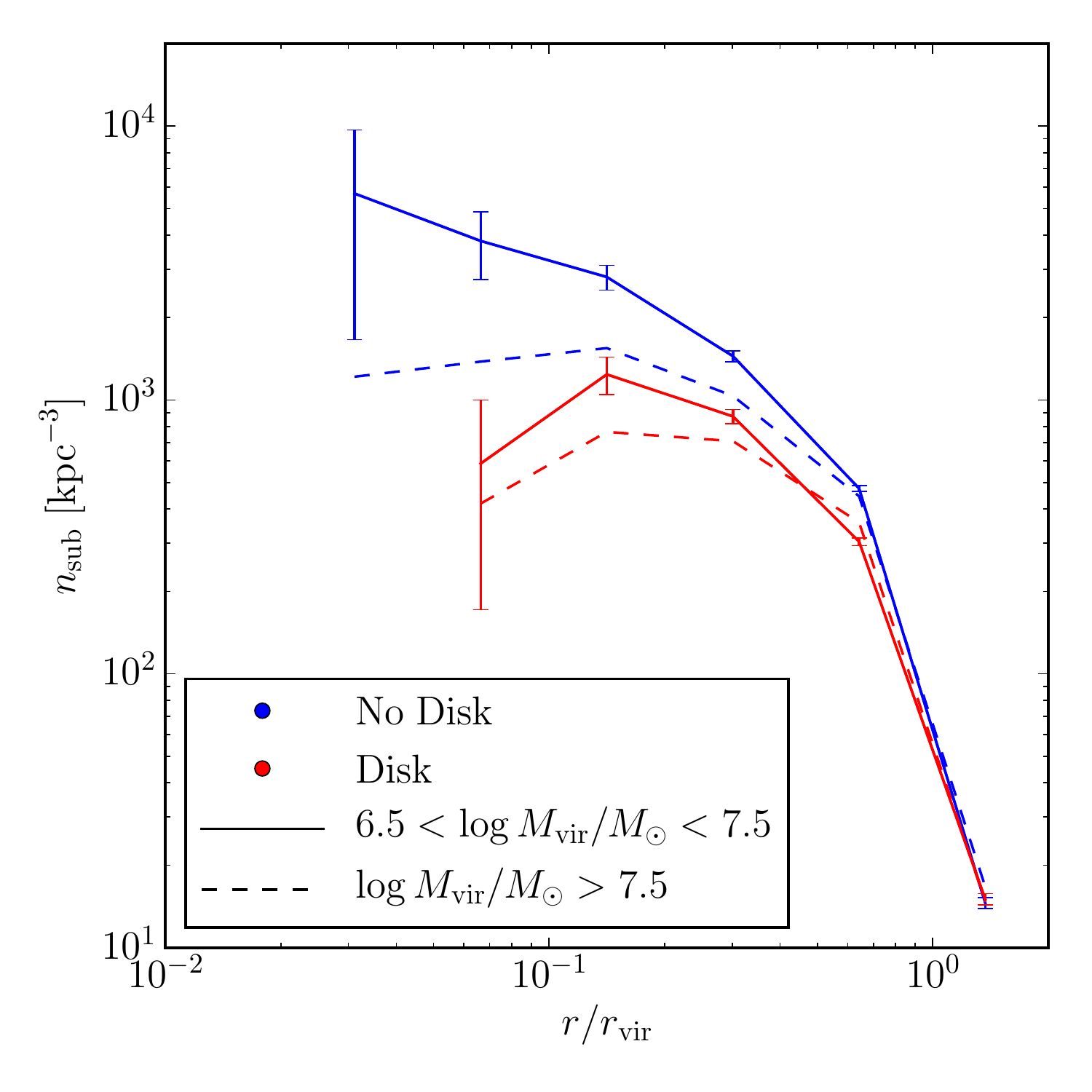}
\caption{
Radial profiles of subhalo number density for simulations with and without a disk, in two different mass bins.
Error bars are Poisson errors on the number of subhalos in a given radial bin, and are a similar size for the high mass bin.
}
\label{fig:radial_prof}
\end{figure}

Figure~\ref{fig:radial_prof} compares the radial profile of subhalo number density for simulations with (red) and without (blue) disks.
This is shown for mass bins $6.5<\log M/M_\odot<7.5$ and $\log M/M_\odot>7.5$, where the low mass bin uses just the high-resolution simulation, while the high mass bin averages over all $N$-body runs.
The destruction of subhalos by the baryonic disk results in a much flatter radial profile within the virial radius.

\subsection{Choice of Stellar Mass Proxy}
\label{ssec:AM_proxy}

In order to assign stellar mass to dark matter halos, we first must decide which halo properties we wish to relate to stellar mass.
\citet{reddick13} investigated the difference between using halo virial mass, $M_\mathrm{vir}$, or maximum circular velocity, $v_\mathrm{max}$, defined at either $z=z_\mathrm{peak}$ (the redshift maximising these quantities), $z=z_\mathrm{acc}$ (the redshift of accretion), or $z=0$.
They found that $v_\mathrm{acc}$ or $v_\mathrm{peak}$ were best able to reproduce clustering statistics when modelling a sample of SDSS galaxies with $M_*>10^{10}M_\odot$.
More recently, \citet{lehmann15} have shown that stellar content likely depends not just on halo mass but also concentration, while \citet{hearin16} have developed a flexible scheme for incorporating dependencies on multiple halo properties in the context of halo occupation distribution models.

The question of which, or how many, halo properties are likely connected to stellar mass are currently restricted to galaxy samples much larger than the few dozen we are dealing with in the present work.
For our purposes, we choose to simply adopt the peak virial mass of a subhalo as the quantity we will relate to its stellar mass.
We choose the peak value because a subhalo's dark matter is more readily stripped than the galaxy it hosts.
We choose $M_\mathrm{vir}$, rather than $v_\mathrm{max}$, since this allows us to more directly constrain the mass of a given satellite galaxy.
All subsequesnt references to $M_\mathrm{vir}$, halo mass, or subhalo mass will refer to peak virial mass unless otherwise specified.

\subsection{Assigning Stellar Mass to Subhalos}
\label{ssec:mstarmods}

We define SMHM relations of the form $P_\mathbf{\Theta}(M_*|M_\mathrm{vir})$, i.e. a probability distribution function that a subhalo with virial mass $M_\mathrm{vir}$ will host a satellite galaxy of stellar mass $M_*$, dependent on model parameters $\Theta$.
We try five models as described below.
A summary of the models and their parameters can be found in Table~\ref{tab:models}.

\begin{table*}
	\centering
	\caption{
	Summary of our SMHM models and, below, their parameters. Prior probability distributions are either uniform in the given range -- i.e. $\mathcal{U}(a,b)$ -- or normally distributed with the given mean and variance -- i.e. $\mathcal{N}(\mu,\sigma^2)$.
	}
	\label{tab:models}
	\begin{tabular}{llll}
	\multicolumn{2}{l}{Model} 								&	Parameters 											& Description																								\\
	\hline
	\multicolumn{2}{l}{Power Law}							&	$M_{11},\alpha$										& power law SMHM with un-informative priors																	\\	
	\multicolumn{2}{l}{Fiducial} 							&	$M_{11},\alpha$										& power law SMHM with informative prior on $M_{11}$															\\
	\multicolumn{2}{l}{Scatter} 							&	$M_{11},\alpha,\sigma_8$							& log-normal scatter in SMHM, increasing for low mass halos 												\\
	\multicolumn{2}{l}{Halo Occupation (HO)}				&	$M_{11},\alpha,M_\mathrm{HO},f_8$					& some halos host no galaxy; occupied halos obey fiducial SMHM 											\\
	\multicolumn{2}{l}{HO + Scatter} 						&	$M_{11},\alpha,M_\mathrm{HO},f_8,\sigma_8$			& HO model with scatter																						\\
	\multicolumn{2}{l}{HO + Broken Power Law} 				&	$M_{11},\alpha,M_\mathrm{HO},f_8,M_T,\beta$			& HO model with occupied halos obeying a double power law SMHM												\\
	\hline
	Parameter 							&	\multicolumn{2}{l}{Description}											& Prior		\\
	\hline
	\multirow{2}{*}{$M_{11}$}			&	\multicolumn{2}{l}{\multirow{2}{*}{stellar mass at $M_\mathrm{vir}=10^{11}M_\odot$}}		& Un-informative: $\log M_{11} \sim \mathcal{U}(5,11)$										\\
	&&& Fiducial: $\log M_{11} \sim \mathcal{N}(8.6,0.4^2)$ i.e. matches other constraints		\\
	$\alpha$							&	\multicolumn{2}{l}{power law slope}															& $\arctan \alpha \sim \mathcal{U}(0,1.33)$	; i.e.	uninformative prior on slope $0\leq\alpha\leq4$	\\
	$\sigma_8$							&	\multicolumn{2}{l}{one-sigma scatter at $M_\mathrm{vir}=10^{8}M_\odot$}						& $\sigma_8 \sim \mathcal{U}(0.2,2)$													\\
	$M_\mathrm{HO}$						&	\multicolumn{2}{l}{mass below which occupation probability < 1}								& $\log M_\mathrm{HO} \sim \mathcal{U}(8.5,10)$											\\
	$f_8$								&	\multicolumn{2}{l}{fraction of occupied $M_\mathrm{vir}=10^{8}M_\odot$ halos} 				& $\log f_8 \sim \mathcal{U}(-3,-1)$													\\
	$M_T$								&	\multicolumn{2}{l}{broken power law break position}											& $\log M_T \sim \mathcal{U}(7,10)$														\\
	$\beta$								&	\multicolumn{2}{l}{broken power law faint end slope}  										& $\arctan \beta \sim \mathcal{U}(0,1.11)$ ; i.e.	$0\leq\beta\leq3$					\\	
	\end{tabular}
\end{table*}

\subsubsection{Power Law/Fiducial}

For our simplest model we assert a one-to-one mapping between stellar mass and halo mass,
\begin{align}
M_*	&= \mu(M_\mathrm{vir}), \nonumber \\
	&:= \min \left[ M_{11} \left( \frac{M_\mathrm{vir}}{10^{11}M_\odot}\right)^\alpha \; , \; f_b M_\mathrm{vir} \right],
\label{eqn:mu}
\end{align}
i.e. a power-law with slope $\alpha$ normalized such that a $10^{11}M_\odot$ halo has $M_*=M_{11}$, with a cap on the maximum allowed stellar mass equal to some fraction $f_b$ of the halo mass.
In the probabilistic framework, the stellar mass pdf is given by a delta function 
\begin{equation}
P_\mathbf{\Theta}(M_*|M_\mathrm{vir}) = \delta (M_* - \mu(M_\mathrm{vir})).
\label{eqn:fid}
\end{equation}

Although $f_b$ could naively be as high as the universal baryon fraction $\Omega_b/\Omega_m=0.16$, we use a lower value of $f_b=0.1$ since we do not expect all baryons to convert into stars.
Though this choice is somewhat arbitrary, our main aim is to place an upper bound on halo mass for a given stellar mass, whereas $f_b$ determines the lower bound, hence our key results do not change when this parameter is increased.

This model has two free parameters: $(M_{11},\alpha)$.
We investigate slopes $0<\alpha<4$, with an uninformative prior which is uniform in $\arctan \alpha$.
We try two different prior distributions on $M_{11}$: the first is un-informative (uniform in $\log$ in the range $10^5 \leq M_{11}/M_\odot \leq 10^{11}$) while the second results from studies of the SMHM relation constrained with stellar mass functions of $10^7 \leq M_*/M_\odot \leq 10^{12}$ galaxies.
We calculate the expected distribution of stellar masses of an $M_\mathrm{vir}=10^{11}M_\odot$ halo assuming the SMHM relations of \citet{behroozi13} and \citet{moster13}.
The two studies agree very well, giving a mean of $\log M_*/M_\odot= 8.6$ and variance of $(0.4$ dex$)^2$.
Our informative prior is given by a normal distribution with these parameters.
Note that this variance is the sum in quadrature of three comparably sized components: intrinsic scatter in the SMHM models, uncertainty in the fit parameters, and the spread arising from the spread in redshifts when halos are accreted.
A power law SMHM relation with this informative prior at $M_\mathrm{vir}=10^{11}M_\odot$ will be henceforth referred to as our fiducial model.

\subsubsection{Scatter}
\label{sssec:scatter}

Introducing scatter in the SMHM relation results in an inferred median SMHM relation which is steeper than would be inferred in the no-scatter case \citep{garrison17}.
This is the result of an Eddington bias: due to the rising subhalo mass function in $Λ$CDM, more low-mass subhalos will be up-scattered to a given stellar mass than high mass subhalos down-scattered.
We also include an SMHM model with scatter, introducing a spread in the permitted $M_*$ at fixed $M_\mathrm{vir}$ given by a log-normal distribution with mean as given by the fiducial model, and a mass-dependent amount of scatter, $\sigma(M_\mathrm{vir})$.

For halo masses $M_\mathrm{vir}>10^{11}M_\odot$ we fix $\sigma=0.2$ dex, in accordance with limits imposed by galaxy abundance and clustering statistics in cosmological volumes \citep[e.g.][]{leauthaud12,reddick13,lehmann15}.
Below $10^{11}M_\odot$, we allow the scatter to grow linearly in $\log M_\mathrm{vir}$,
\begin{equation}
\sigma(M_\mathrm{vir}) = 
\begin{cases}
	0.2 \qquad\qquad\qquad\qquad\quad \text{if} \ M_\mathrm{vir}> 10^{11} M_\odot,\\
	\dfrac{0.2-\sigma_8}{3}\left(\log \dfrac{M_\mathrm{vir}}{M_\odot} - 8\right) + \sigma_8 \quad \text{otherwise,}
\end{cases}
\label{eqn:sigma}
\end{equation}
where we have parameterized its growth using the value $\sigma_8=\sigma(10^8M_\odot)$, not to be confused with the $\sigma_8$ from large scale structure.
As with the fiducial model, we cap the maximum allowed stellar mass as a fraction $f_b=0.1$ of the halo mass, so that the total model is given by
\begin{equation}
P_\mathbf{\Theta}(\log M_*|\log M_\mathrm{vir}) = 
\begin{cases}
K \mathcal{N}(\log \mu, \sigma^2) 	& \text{if} \; M_* <  f_b M_\mathrm{vir},\\
0									& \text{otherwise,}
\end{cases}
\label{eqn:scatter}
\end{equation}
where $\mathcal{N}$ is a normal distribution parametrised by its mean and variance, $\mu$ is defined in Equation~(\ref{eqn:mu}), $\sigma$ from Equation~(\ref{eqn:sigma}) and $K$ is a normalizing constant.

This model has three free parameters: $(M_{11},\alpha,\sigma_8)$.
Priors on $M_{11}$ and $\alpha$ are as for the fiducial model, while we investigate values $0.2 \leq \sigma_8 / \mathrm{dex} \leq 2$ with a prior flat in this range.
Note that the edge value of $\sigma_8=0.2$ would give a constant scatter of 0.2 dex at all masses.

\subsubsection{Halo Occupation}

Our next class of model allows halos below some threshold mass to not host any galaxy.
This idea is motivated by calculations which suggest that the energetic UV background created during the epoch of reionisation will efficiently dissociate H$_2$ in prospective dwarf galaxies, with the loss of this efficient coolant resulting in gas supplies too warm to remain bound to halos below some threshold mass \citep[e.g.][]{ikeuchi86,rees86,efstathiou92}.

We explore the results of two sets of simulations to decide how best to incorporate this effect into our model.
Firstly, we consider \citet{okamoto08}.
They performed cosmological, hydrodynamic simulations of galaxy formation with a time evolving UV background, quantifying the characteristic mass at which a halo (on average) loses half of its baryons to photo-heating.
Interestingly, they find that this characteristic mass is much smaller than the filtering mass predicted by linear perturbation theory \citep{gnedin98,gnedin00} and, furthermore, it evolves from $10^8M_\odot$ at $z=6$ to $\sim5\times10^9M_\odot$ at $z=0$, subject to their assumed thermal history.
There is a spread $\sim1$ dex about the characteristic mass where baryon fraction drops from the cosmological value to 0, and in this range galaxy formation may become inefficient, but not necessarily shut down completely.

Secondly, we consider \citet{sawala16}, whose hydrodynamic, cosmological simulations additionally include sub-grid physics modules \citep{schaye15,crain15} which allow for treatment of physical processes occurring below the resolution limit, e.g. star formation.
They quantify the occupancy fraction of their simulated halos as a function of halo mass, where an occupied halo contains a luminous galaxy, i.e. at least one star particle.
Above some redshift dependent threshold mass, they find that all halos are occupied.
Subject to their assumed thermal history, this threshold evolves from $3\times10^8M_\odot$ at $z=11.5$ to $10^{10}M_\odot$ at $z=0$.
At all epochs, they find that the fraction of $10^{8}M_\odot$ halos which host a galaxy remains below 10\%.

To incorporate this variable occupancy into our model, we define a halo occupation probability $p_\mathrm{HO}$ with a mass dependence described by a power-law decay below some threshold mass $M_\mathrm{HO}$,
\begin{equation}
p_\mathrm{HO}(M_\mathrm{vir}) = 
\begin{cases}
1										 							& \text{if} \; M_\mathrm{vir}>M_\mathrm{HO},\\
\left(\dfrac{M_\mathrm{vir}}{M_\mathrm{HO}}\right)^\gamma			& \text{otherwise.}
\end{cases}
\label{eqn:occfrac}
\end{equation}
For ease of interpretation, we parametrise the exponent $\gamma$ as
\begin{equation}
\gamma = \dfrac{\log f_8}{8 - \log M_\mathrm{HO}/M_\odot}
\label{eqn:gamma}
\end{equation}
such that $f_8$ is the fraction of $10^{8}M_\odot$ subhalos which are occupied.
The SMHM model is then given by
\begin{equation}
P_\mathbf{\Theta}(M_*| M_\mathrm{vir}) = (1-p_\mathrm{HO}) \delta(M_*) + p_\mathrm{HO} \delta(M_* - \mu(M_\mathrm{vir}))
\label{eqn:haloocc}
\end{equation}
where the $\delta$ are delta functions and $\mu$ is the power-law relation defined in Equation~(\ref{eqn:mu}).

This model has four free parameters: $(M_{11},\alpha,M_\mathrm{HO},f_8)$.
Priors on $M_{11}$ and $\alpha$ are as for the fiducial model.
Guided by the evolution of threshold masses in \citet{okamoto08} and \citet{sawala16}, we expect strong evolution of $M_\mathrm{HO}$ with redshift.
Given that our model assigns stellar mass based only on peak halo mass, however, we also evaluate occupation probability based on peak halo mass, rather than use a more complicated model of the form $p_\mathrm{HO}(M,z)$.
By using a broad prior on the value of $M_\mathrm{HO}$, however, we hope to gain some insight into the evolution of the threshold mass and its link to reionisation.
Our prior on $M_\mathrm{HO}$ is uniform in $\log$ in the range $8.5 \leq \log M_\mathrm{HO}/M_\odot \leq 10$, spanning the evolution between $0<z<z_\mathrm{rei}$ of the threshold mass in \citet{sawala16}.
Our prior on $f_8$ is uniform in $\log$ in the range $10^{-3}<f_8<10^{-1}$.
This upper bound corresponds to the 10\% cap on occupation fraction at $M_\mathrm{vir}=10^8M_\odot$ observed in \citet{sawala16}, while the lower bound of $f_8=0.001$ results in an expected number of fewer than 0.1 occupied $10^8M_\odot$ subhalos inside the virial radius.

\subsubsection{Halo Occupation and Scatter}

This is an extension of the halo occupation model where we also include a log normal scatter in stellar mass in the same way as described in Section~\ref{sssec:scatter}.
There are five free parameters, $(M_{11},\alpha,\sigma_8,M_\mathrm{HO},f_8)$, with priors as discussed previously.

\subsubsection{Halo Occupation and Broken Power Law}
\label{ref:}

As a final model, we consider an extension of the halo occupation model where we replace the fiducial SMHM relation with a broken power law.
This is in part motivated by the \citet{okamoto08} simulations which, as discussed earlier, predict that around the characteristic mass galaxy formation may become inefficient, leading to a steepening of the SMHM relation.
On the other hand, \citet{sawala15} predict that the SMHM becomes shallower at low masses.
We use a broken power law model, able to accommodate both of these alternatives, given by,
\begin{align}
M_* &= \nu(M_\mathrm{vir}), \nonumber \\
	&:=
		\begin{cases}
		\mu(M_\mathrm{vir})						 																	& \text{if} \; M_\mathrm{vir}>M_T,\\
		\min \left[ \mu(M_T) \left(\dfrac{M_\mathrm{vir}}{M_T}\right)^\beta \; , \; f_b M_\mathrm{vir} \right]		& \text{otherwise.}
		\end{cases}
\label{eqn:bpl}
\end{align}
which has a SMHM relation with slope $\beta$ below a transition mass $M_T$.
The new model is given by replacing $\mu$ with $\nu$ in Equation~(\ref{eqn:haloocc}).
This model has six free parameters, $(M_{11},\alpha,M_\mathrm{HO},f_8,M_T,\beta)$.
The priors on the first four are as before.
Our prior on $M_T$ is uniform in $\log$ in the range $7 \leq \log M_T/M_\odot \leq 9$.
We investigate faint-end slopes $0<\beta<3$, with an uninformative prior which is uniform in $\arctan \beta$.

\subsection{Data and Selection Function}
\label{ssec:data}

\begin{table}
	\centering{}
	\caption{
	The names, $V$-band absolute magnitude, stellar mass and heliocentric distance of the satellite galaxies used for this analysis.
	Magnitudes and distances are taken from \citet{mcconnachie12}, stellar masses from \citet{woo08} where available, otherwise assuming a stellar mass to light ratio $M_*/L=2M_\odot/L_\odot$.
	}
	\label{tab:data}
	\begin{tabular}{lccc}
	Name 	& 	$M_V$ (mag)	&	$M_*$ ($M_\odot$) 	&	$D_\odot$ (kpc) \\
	\hline
	\hline
	LMC                 &   -18.1   &   $1.1\times10^9$   	&   51.0 \\
	SMC                 &   -16.8   &   $3.7\times10^8$		&   64.0 \\
	Sagittarius         &   -13.5   &   $3.4\times10^7$		&   26.0 \\
	Fornax              &   -13.4   &   $3.3\times10^7$		&   147. \\
	Leo I               &   -12.0   &   $8.8\times10^6$		&   254. \\
	Sculptor            &   -11.1   &   $3.7\times10^6$		&   86.0 \\
	Leo II              &   -9.8    &   $1.2\times10^6$		&   233. \\
	Sextans I           &   -9.3    &   $7.0\times10^5$		&   86.0 \\
	Carina              &   -9.1    &   $6.0\times10^5$		&   105. \\
	Draco               &   -8.8    &   $4.5\times10^5$		&   76.0 \\
	Ursa Minor          &   -8.8    &   $4.5\times10^5$		&   76.0 \\
	Canes Venatici I    &   -8.6    &   $1.2\times10^5$		&   218. \\
	Hercules            &   -6.6    &   $1.9\times10^4$		&   132. \\
	Bootes I            &   -6.3    &   $1.4\times10^4$		&   66.0 \\
	Leo IV              &   -5.8    &   $9.0\times10^3$		&   154. \\
	Ursa Major I        &   -5.5    &   $6.8\times10^3$		&   97.0 \\
	Leo V               &   -5.2    &   $5.1\times10^3$		&   178. \\
	Pisces II           &   -5.0    &   $4.3\times10^3$		&   182. \\
	Canes Venatici II   &   -4.9    &   $3.9\times10^3$		&   160. \\
	Ursa Major II       &   -4.2    &   $2.1\times10^3$		&   32.0 \\
	Coma Berenices      &   -4.1    &   $1.9\times10^3$		&   44.0 \\
	Bootes II           &   -2.7    &   $5.1\times10^2$		&   42.0 \\
	Willman 1           &   -2.7    &   $5.1\times10^2$		&   38.0 \\
	Segue II            &   -2.5    &   $4.3\times10^2$		&   35.0 \\
	Segue I             &   -1.5    &   $1.7\times10^2$		&   23.0 \\
	\end{tabular}
\end{table}

The data we will use to constrain our model are the luminosities of the 11 brightest, classical, MW satellite galaxies and fainter galaxies discovered in Sloan Digital Sky Survey (SDSS) data releases up to and including DR9 \citep{ahn12}.
These are listed in Table~\ref{tab:data} along with relevant galaxy properties.
Though numerous satellite galaxies have been discovered in a number of surveys subsequent to SDSS, we choose to exclude these from our analysis for two reasons.

First, of the $\sim20$ new discoveries recently made in in the Dark Energy Survey (DES) \citep{koposov15,DES_Y1,drlicawagner15,kim15a,kim15b}, the majority are most likely associated with the LMC \citep{jethwa16}.
Since our zoom-in simulations were not explicitly chosen to host a subhalo as massive as the LMC's \citep[$M_\mathrm{vir}\sim10^{11}M_\odot$, from e.g.][]{moster13,penarrubia16}, and such massive subhalos are rare around MWs in $\Lambda$CDM \citep{boylankolchin11}, it would not be a fair comparison to include the DES satellites in this analysis.
It may even be possible that some of the SDSS-DR9 satellites also have an origin in the LMC though this more unlikely, due to the distance between the LMC and the SDSS footprint.
We discuss this further in Section~\ref{sec:discuss}.

Second, we restrict our analysis to SDSS is that it has a known selection function for satellite detection \citet{koposov08} which we will apply to our model to mimic observational biases.
This can be approximated as a threshold magnitude as a function of distance \citep{koposov09}
\begin{equation}
M_V < (1.1 - \log_{10}(D_\odot/\mathrm{kpc}))/0.228,
\label{eqn:selc_func}
\end{equation}
fainter than which a satellite is undetectable.
A further requirement for detectability is that the dwarf exceeds the SDSS surface brightness threshold of 30 mag arcsec$^{-2}$ \citep{koposov08}.
Since our model does not predict the physical extent of a dwarf's stellar component, and hence its surface brightness, we will make the simplifying assumption that all dwarfs that exist pass the surface brightness requirement.
We will discuss the implication of this assumption in Section~\ref{sec:discuss}.

We highlight that the faintest galaxy in this sample is the $M_V=-1.5$ Segue I.
The classification of Segue I as either a dwarf galaxy or star cluster was originally uncertain due to foreground contamination in estimates of its dynamical mass \citep{niedersteostholt09}.
Using a Bayesian treatment of membership probabilities, however, \citet{martinez11} showed that is is most likely a dark matter dominated dwarf galaxy and hence we include it in our sample.

\subsection{Probabilistic Model}
\label{ssec:probmod}

We now describe the steps taken to calculate the likelihood $P(\mathbf{X}|\Theta)$, i.e. the probability of observing the MW satellite luminosities, $\mathbf{X}$, given a particular SMHM model and its parameters, $\Theta$.
We first generate a set of model satellite luminosities, $\mathbf{M}$.
To do this, for a given $N$-body simulation, we assign stellar mass to subhalos according to the SMHM model and parameters, $\Theta$.
We then mimic the observational biases which have gone in to the galaxy sample described in Section~\ref{ssec:data}.
For model satellite galaxies fainter than $M_V=-11$ we observe only those within the SDSS-DR9 footprint brighter than the detection threshold given by Equation~\ref{eqn:selc_func}.
Satellites brighter than $M_V=-11$, corresponding to the classical MW satellites, are observed over the entire sky.

We now wish to compare the discrete observed and simulated samples, $\mathbf{X}$ and $\mathbf{M}$.
\citet{garrison17} do this by defining a metric based on the difference between the rank-ordered data and model luminosities.
Though this suffices to differentiate good and bad models, such an approach prohibits a probabilistic inference on the SMHM relation.
To facilitate the required calculation of $P(\mathbf{X}|\mathbf{M})$, we introduce luminosity bins $i\in\{1,...,N_\mathrm{bins} \}$ which are chosen to minimize the bias and variance of inferences performed with fake data as discussed in Section~\ref{ssec:fakedata}.

Assuming that the number of data and model satellites in bin $i$ is $X_i$ and $M_i$ respectively, and assuming independent bins, gives a probability,
\begin{equation}
P(\mathbf{X}|\mathbf{M}) = \prod_{i=1}^{N_\mathrm{bins}} P(X_i|M_i).
\label{eqn:p_x_m}
\end{equation}
The assumption of independent bins is technically invalid, however we show in Section~\ref{ssec:fakedata} that this negligibly affects our results.
Assuming small numbers in each bin, a reasonable guess for $P(X_i|M_i)$ would be a Poisson distribution, $P_{M_i}(X_i)$, where
\begin{equation}
P_\lambda(n) = \dfrac{\lambda^n e^{-\lambda}}{n!}.
\label{eqn:poiss}
\end{equation}
However, we considered this unsatisfactory as it does not account for the stochasticity within the subhalo distribution in $N$-body simulations.
To illustrate this, imagine that one bin contains only the MW's most massive satellite, the LMC.
A Poisson distribution would assign a probability of zero to any model predicting zero satellites in this bin, even if that model well reproduced the luminosity function for fainter satellites.
The Poisson probability does not take into account the inherent randomness in whether or not a simulated halo will host a subhalo massive enough to host the LMC.

We relax this rigidity in the likelihood by setting the mean Poisson rate of model satellites in bin $i$ equal not to $M_i$, but instead to some unknown variable $\lambda_i$.
The $X_i$ data- and $M_i$ model satellites are then draws from a Poisson process with mean $\lambda_i$, i.e.
\begin{align}
P(X_i|\lambda_i) = P_{\lambda_i}(X_i), 
\label{eqn:p_xi_li}
\\
P(M_i|\lambda_i) = P_{\lambda_i}(M_i)
\label{eqn:p_mi_li}
\end{align}
The probability $P(X_i|M_i)$ is found by marginalising over all possible values of $\lambda_i$, 
\begin{align}
P(X_i|M_i)	&=\int P(X_i|\lambda_i) P(\lambda_i|M_i) \; \mathrm{d}\lambda_i,	\\
			&\propto\int P(X_i|\lambda_i) P(M_i|\lambda_i) P(\lambda_i) \; \mathrm{d}\lambda_i,
\label{eqn:p_xi_mi_0}
\end{align}
where to arrive at the second line we use Bayes' theorem and introduce the prior probability $P(\lambda_i)$.
Inserting the Poisson probabilities of Equations~(\ref{eqn:p_xi_li}) and~(\ref{eqn:p_mi_li}), and assuming $P(\lambda_i)$ is a constant for $\lambda_i \geq 0$, the above integral evaluates to
\begin{equation}
P(X_i|M_i) = 2^{-(X_i+M_i+1)}\dfrac{(X_i+M_i)!}{X_i!M_i!}
\label{eqn:p_xi_mi}
\end{equation}

Inserting Equation~(\ref{eqn:p_xi_mi}) into~(\ref{eqn:p_x_m}) defines our likelihood for a given sample of model luminosities $\mathbf{M}$.
For a given $N$-body simulation, we can draw multiple samples $\mathbf{M}$ by randomizing various nuisance parameters.
Firstly, we vary the position of the Sun, allowing it to lie anywhere on the sphere with galactocentric radius $R_\odot=8.3$ kpc \citep{mcmillan11} for the no-disk simulations, and restricting this to the disk mid-plane for the simulations with a disk.
Secondly, for scatter and halo occupancy models, we also randomize the stellar mass of a given subhalo.
For each simulation, we take $N_\mathrm{rand}=30$ random draws -- at which value our results are converged -- with the total likelihood for a simulation equal to the average of the $N_\mathrm{rand}$ lots of $P(\mathbf{X}|\mathbf{M}) $.
Finally, we average the likelihood over a set of simulations.
This will typically be all of those either with or without a disk.
In full, and now including the implicit dependence of model satellite luminosities $\mathbf{M}$ on simulation, nuisance parameters and model, we have,
\begin{equation}
P(\mathbf{X}|\Theta) = \dfrac{1}{|S|} \dfrac{1}{N_\mathrm{rand}} \sum_{\mathrm{sim} \in S} \sum_{j=1}^{N_\mathrm{rand}} 
P(\mathbf{X}|\mathbf{M}(\mathrm{sim},j,\Theta)),
\label{eqn:p_x_theta}
\end{equation}
where $S$ is the set of simulations.

\subsection{Fake Data Tests \& Bin Selection}
\label{ssec:fakedata}

To facilitate the comparison of the discrete model- and data- satellite luminosities, we have introduced luminosity bins and constructed a likelihood function under the assumption of bin independence.
However, this assumption is wrong: a model which predicts many satellites in one bin is more likely to predict many satellites in a neighboring bin than some competing model.
Therefore, though finer binning retains more information and hence may decrease the error due to bias in our inference, this comes at the expense of artificially increasing the number of degrees of freedom in the fit, which may increase the error due to variance.
Using fake data tests, we now look for the optimal binning to trade-off between bias and variance, and minimize the total error in our inference.

From our fiducial model of the SMHM relation, we sample five pairs of model parameters $(M_{11}/M_\odot,\alpha) \in \{(10^{10},3.5),(10^{10},2),(10^9,3),(10^9,2.3),(10^{7.5},1.8)\}$, and for each pair generate a fake data set from each of the seven disk-free medium resolution $N$-body simulations, totalling 35 fake data sets.
We fit each one using the fiducial SMHM model, using the probabilistic framework described in the previous section, excluding from the fit the simulation from which the fake data was generated.
We test six choices of binning, letting $N_\mathrm{bins} \in \{1,0.25 N_\mathrm{dat}, 0.5 N_\mathrm{dat}, 0.75 N_\mathrm{dat}, N_\mathrm{dat}, 2N_\mathrm{dat} \}$ where $N_\mathrm{dat}$ is the size of the fake data set and fractional values are rounded to the nearest integer.
The bins are spaced evenly in absolute magnitude between the faintest and brightest satellites in the sample.

To evaluate fit performance we calculate the error on our inference on the halo mass of the faintest satellite in the fake data set, whose stellar mass we denote $M_*^\mathrm{min}$.
Ignoring any irreducible error due to data or systematics, this is given by the expected value of the squared difference between our inferred halo mass $M_\mathrm{vir}$ and the true value $M_\mathrm{vir}^\mathrm{min}$,
\begin{equation}
\mathrm{Error}^2 = \mathbb{E}[(M_\mathrm{vir}-M_\mathrm{vir}^\mathrm{min})^2],
\label{eqn:error}
\end{equation}
where $M_\mathrm{vir}^\mathrm{min} = \mu^{-1}(M_*^\mathrm{min})$ and $\mu$ is the fiducial SMHM relation in Equation~(\ref{eqn:mu}).
This expectation is probability-weighted by the inferred posterior predictive pdfs on the halo mass at fixed stellar mass of $M_*^\mathrm{min}$ (see Section~\ref{ssec:invert_constraints} for details).
The error can be decomposed as
\begin{equation}
\mathrm{Error}^2= (\mathbb{E}[M_\mathrm{vir}-M_\mathrm{vir}^\mathrm{min}])^2 + (\mathbb{E}[M_\mathrm{vir}^2]-\mathbb{E}[M_\mathrm{vir}]^2) , \nonumber \\
\label{eqn:bias_var}
\end{equation}
where the first term is the square of the bias, and the second is the variance.

\begin{figure}
\includegraphics[width=\columnwidth]{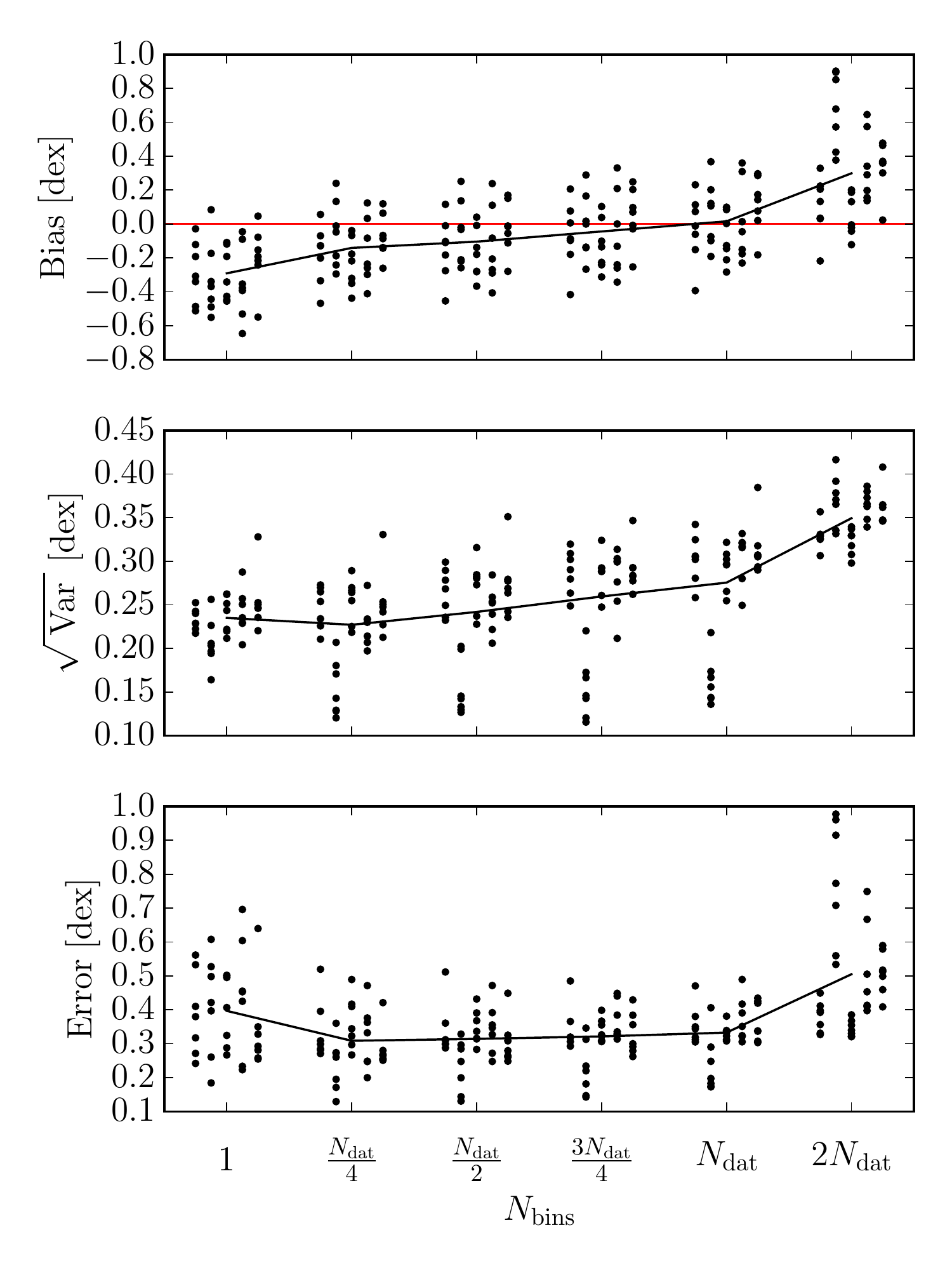}
\caption{
Error in our inference of the halo mass of the faintest satellite in fake data tests, as a function of the number of bins used.
Panels show the bias (top), variance (middle) and total error (bottom) of the inference for individual fake data sets (circles) and the mean for each value of $N_\mathrm{bins}$ (black lines).
The red line in the top panel highlights zero bias.
The 5 horizontally offset columns shown for each value of $N_\mathrm{bins}$ correspond to the five sets of the model parameters $(M_{11}/M_\odot,\alpha)$ used to generate the fake data.
}
\label{fig:bias_var}
\end{figure}

Figure~\ref{fig:bias_var} shows the bias, variance and total error as a function of the number of bins used.
As we increase the number of bins used from 1 to $N_\mathrm{dat}$, the absolute value of the mean bias shrinks from -0.3 to 0.01 dex, at the expense of variance increasing from 0.23 to 0.28 dex.
Increasing $N_\mathrm{bins}$ beyond $N_\mathrm{dat}$ leads to an increase of the bias and variance.
The total error roughly plateaus in the range $N_\mathrm{dat}/4 \leq N_\mathrm{bins} \leq N_\mathrm{dat}$, and is formally minimized at $N_\mathrm{bins}=N_\mathrm{dat}/4$.
We take this value for all further analysis.
For our real data set with $N_\mathrm{dat}=25$, this results in 6 bins evenly spaced in the range $-20<M_V<-1$, i.e. the range of our data.

\section{Comparison of SMHM models}
\label{sec:results}

We calculate the posterior pdf on the parameters of our six models of the SMHM relation.
Using Bayes' theorem this is given by,
\begin{equation}
P(\Theta|\mathbf{X}) = \dfrac{P(\mathbf{X}|\Theta)P(\Theta)}{\int P(\mathbf{X}|\Theta)P(\Theta) \; \mathrm{d}\Theta},
\label{eqn:posterior}
\end{equation}
where $P(\mathbf{X}|\Theta)$ is the likelihood from Equation~\ref{eqn:p_x_theta}, $P(\Theta)$ are the parameter priors listed in Table~\ref{tab:models}, and the normalizing denominator is the model evidence.
We calculate the likelihood over a parameter grid with between 10-100 values per parameter (dependent on model dimensionality), sampled uniformly over the prior distributions.
In this section we compare the performance of different SMHM models and models of halo occupation.

\subsection{Bayes Factor}

\begin{table}
	\centering
	\caption{
	Bayes factors comparing SMHM models, calculated in reference to the fiducial model separately for disk/no-disk models.
	}
	\label{tab:bayes_factors}
	\begin{tabular}{lcc}
	SMHM Model 				&	No-disk Sims 			&	Disk Sims 			\\
	\hline
	Power Law 				&	1.37					&	1.52				\\
	Fiducial				&	1.00					&	1.00				\\
	Scatter 				&	1.34					&	1.43				\\
	Halo Occupancy (HO) 	&	2.11					&	3.02				\\
	HO + Scatter 			&	2.56					&	3.77				\\
	HO + Broken Power Law 	& 	2.41					& 	3.32
	\end{tabular}
\end{table}

To compare two models $M_1$ and $M_2$, parameterized by $\Theta_1$ and $\Theta_2$ respectively, we can calculate the Bayes factor $K$, which is the ratio of model evidences
\begin{equation}
K = \dfrac{\int P(\mathbf{X}|M_1,\Theta_1)P(\Theta_1|M_1) \; \mathrm{d}\Theta_1}{\int P(\mathbf{X}|M_2,\Theta_2)P(\Theta_2|M_2) \; \mathrm{d}\Theta_2}.
\label{eqn:bayesf}
\end{equation}
where we have explicitly included the conditioning of the likelihood and prior probabilities on the model in question.
A value of $K>1$ means that $M_1$ is more strongly supported by the data than $M_2$, where the strength of evidence for this claim is ``not worth more than a bare mention'' for values $1<K<3$, ``positive'' for $3<K<20$, and ``strong'' for $K>20$ \citep{Kass95}.
This measure of model goodness naturally accounts for complexity, penalizing models with more parameters than are warranted by the data.

In Table~\ref{tab:bayes_factors} we show Bayes factors for the six models of the SMHM relation.
We calculate these independently for models with and without a disk, in both cases doing this in reference to the fiducial SMHM model, which is the favored model for both cases.
For the simulations with no disk, we see that there is little evidence for any one model's superiority over any other, however all models with limited halo occupation are slightly disfavored.
This result is strengthened when we look at the simulations with a disk, for which there is positive evidence (i.e. $K>3$) that models with halo occupancy fraction limited to 10\% are disfavored by the data.
We investigate this further below.

\subsection{Posterior Predictive Luminosity Functions}
\label{ssec:posterior_predictive}

To see why the data disfavor certain models, we will now show a comparison of data and models.
Rather than showing this at a single point in parameter space (e.g. the maximum likelihood value) we will achieve a more statistically thorough comparison by showing how well a model can reproduce the data when allowed to roam over the entire scope of its parameter space.
In other words, we will calculate the the posterior predictive distribution on the observed luminosity function.

\begin{figure*}
\includegraphics[width=\textwidth]{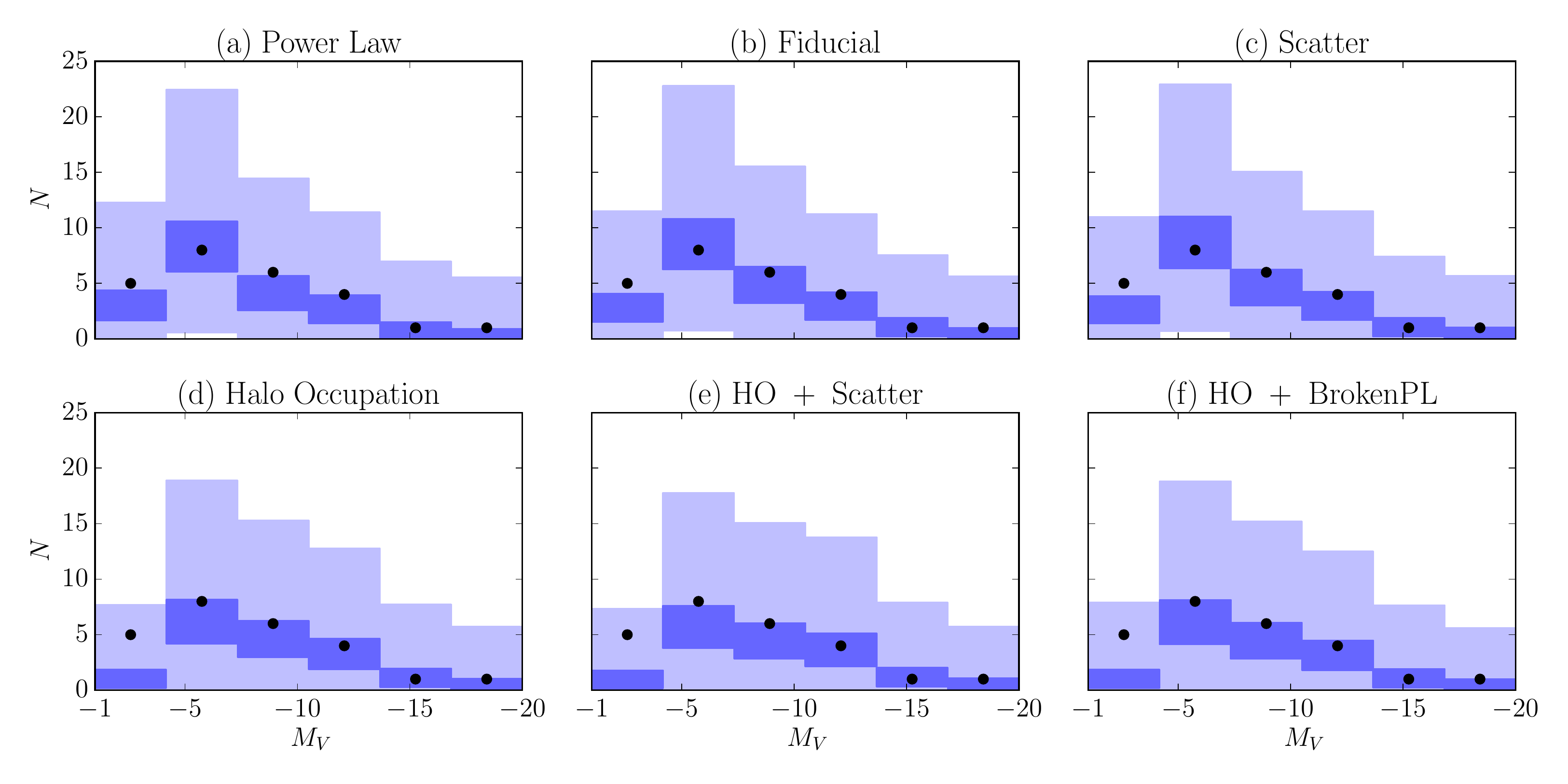}
\caption{
Posterior predictive luminosity functions for simulations with no disk.
These are shown for six different SMHM models as indicated by panel titles.
The dark/light shaded regions show the 68/95\% confidence intervals for the number of observed satellites in each luminosity bin.
The black circles show the observed data.
}
\label{fig:pplf_nodisk}
\includegraphics[width=\textwidth]{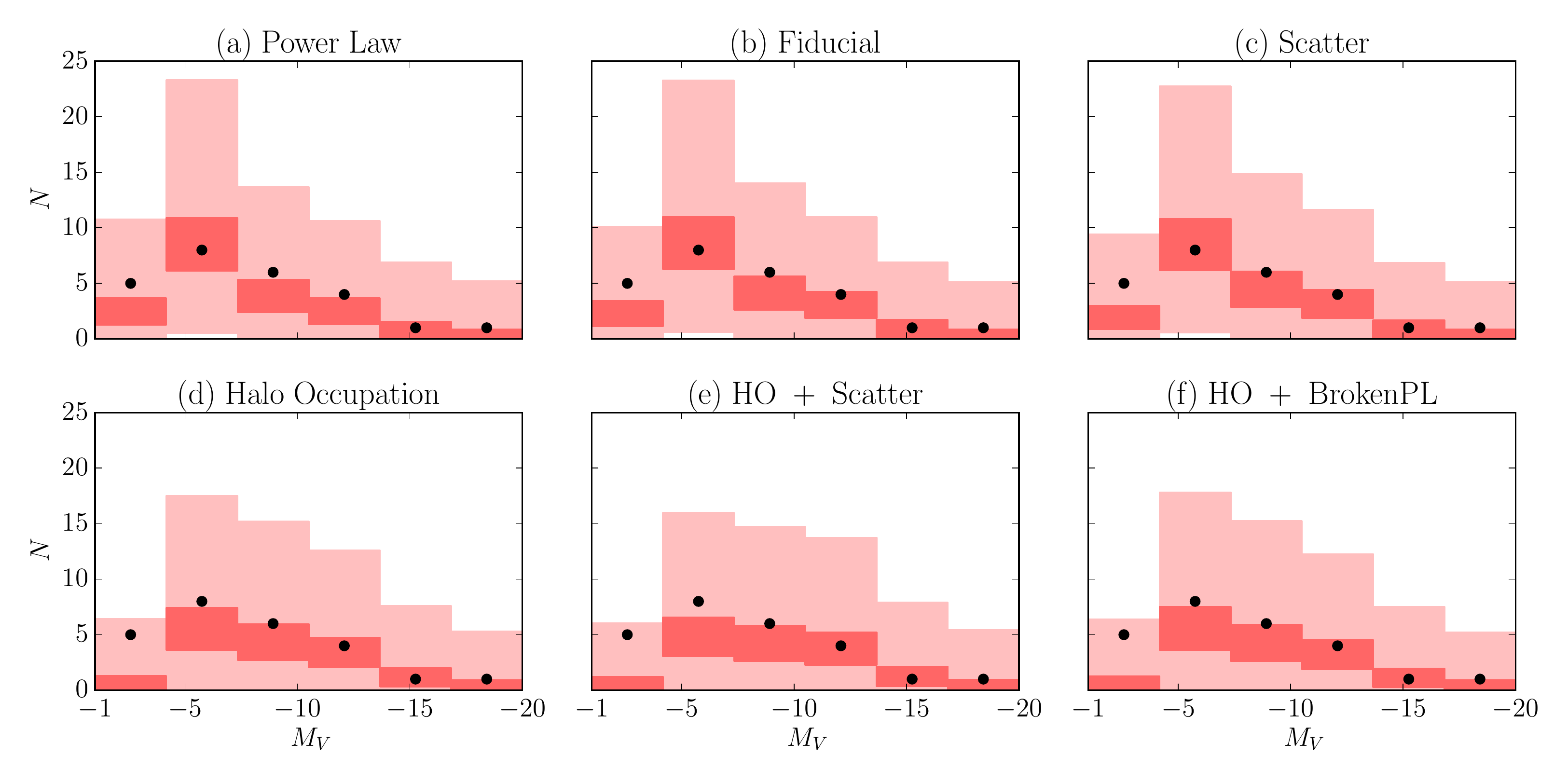}
\caption{As Figure \ref{fig:pplf_nodisk} but for simulations with a disk.}
\label{fig:pplf_disk}
\end{figure*}

Letting $\lambda_i$ be the number of satellites in luminosity bin $i$, the posterior predictive distribution on $\lambda_i$ conditioned on the observed data $\mathbf{X}$ is given by
\begin{align}
P(\lambda_i|\mathbf{X}) = &\dfrac{1}{|S|} \dfrac{1}{N_\mathrm{rand}} \sum_{\mathrm{sim} \in S} \sum_{j=1}^{N_\mathrm{rand}} \nonumber \\
& \int P(\lambda_i|M_i) P(\mathbf{X}|\mathbf{M}(\mathrm{sim},j,\Theta)) P(\Theta) \; \mathrm{d}\Theta
\label{eqn:pp_lumfunc}
\end{align}
where the summations are over different simulations and random draws of nuisance parameters.
All notation is as used in Section~\ref{ssec:probmod}.
Figures \ref{fig:pplf_nodisk} and \ref{fig:pplf_disk} show the posterior predictive luminosity functions for models without and with a disk, respectively.
We now discuss the result of using different SMHM models, and defer discussion of the difference between the simulations with and without a disk to Section~\ref{sec:disk_nodisk}.

\subsubsection{Power Law, Fiducial \& Scatter models: good fits}

The power law, fiducial and scatter SMHM models produce luminosity functions which well reproduce the data for both disk and no-disk cases (see panels a,b and c of Figures \ref{fig:pplf_nodisk} and \ref{fig:pplf_disk}).
In every luminosity bin the observed number of satellites is either consistent with the model predictions at the 1$\sigma$ level, or only marginally larger than the 1$\sigma$ prediction -- this second option always holding true for the faintest luminosity bin.

These three models are able to reproduce the observed data equally well.
This is true despite the power law model exploring a larger parameter space than the fiducial model, and despite the scatter model having an increased dimensionality.
This means that the slight preference we infer for the fiducial model over the other two -- as evidenced by the Bayes factors in Table~\ref{tab:bayes_factors} -- must be driven entirely by their increased complexity, which comes without any discernible improvement in the fit quality.

\subsubsection{Halo Occupancy models: too few faint satellites}
\label{subsubsec:ho_pplf}

\begin{figure*}
\includegraphics[width=\textwidth]{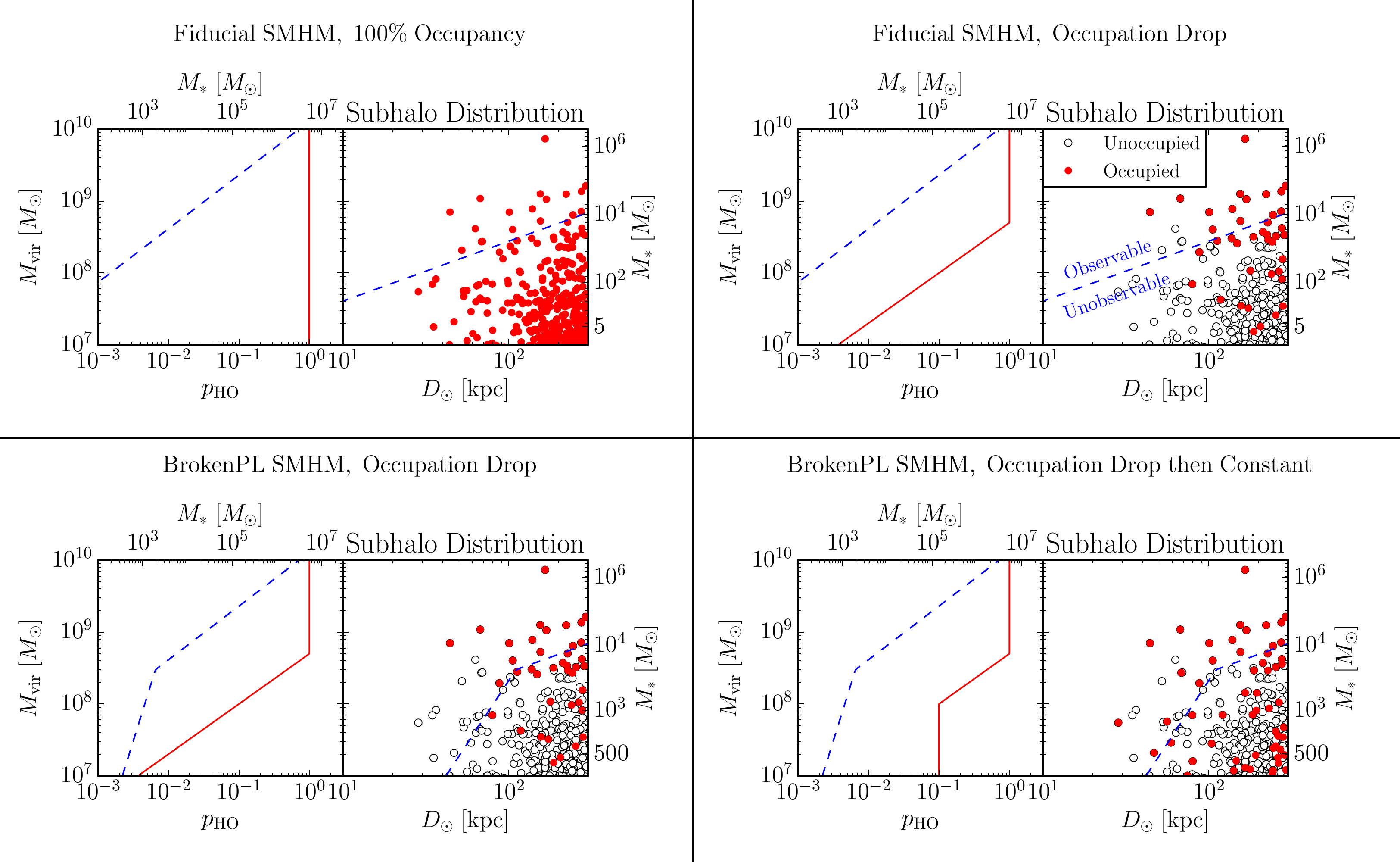}
\caption{
Example distributions of subhalos/satellite galaxies, where each quadrant represents four different models of halo occupation and/or the SMHM relation.
For each quadrant, the left panel shows halo mass ($y$-axis) against the halo occupation probability $p_\mathrm{HO}$ (solid red line) and the SMHM relation (dashed blue line).
Given this model, for one random realization from one of our $N$-body simulations, the right panels shows the distance ($x$-axis) of subhalos which are occupied (red circles) or un-occupied (white circles) by luminous galaxies, and the boundary between observable and un-observable (dashed blue line).
}
\label{fig:halo_occupancy_compare}
\end{figure*}

Restricting the fraction of $10^8M_\odot$ halos which host a galaxy to at most 10\%, we under-predict the observed number of satellites in the faintest luminosity bin.
For the simulations with no disk, the observed number of satellites in this bin lies half way between the $1\sigma$ and $2\sigma$ predictions (see Figure \ref{fig:pplf_nodisk}, panels \textit{d, e} and \textit{f}).
When we include a baryonic disk, the observed count in the faintest bin lies at the edge of the 2$\sigma$ prediction (see Figure \ref{fig:pplf_disk}, panels \textit{d, e} and \textit{f}).
This happens because the disk destroys subhalos, as shown by the reduction in subhalo number density in Figure~\ref{fig:radial_prof}.
Thus there are simply fewer surviving low-mass subhalos able to host ultra-faint satellites in models with a disk, exacerbating the tension between model and data.

Figure~\ref{fig:halo_occupancy_compare} explicitly demonstrates the effect of restricting halo occupation.
Each quadrant represents a different SMHM relation/halo occupation model, and the satellite galaxy distribution which results from applying this to one view of our $N$-body simulations.
The upper-left quadrant shows 100\% halo occupation and a power law SMHM relation with a slope and normalization which give a predicted luminosity function similar to panel \textit{a} of Figure~\ref{fig:pplf_disk}, i.e. a good match to the data.
The upper-right panel has the same SMHM relation, but we decrease the occupation fraction below $M_\mathrm{vir}=5\times10^8M_\odot$, which results in a luminosity function similar to panel \textit{d} of Figure~\ref{fig:pplf_disk}, i.e. with a deficit of faint satellites.
Though Figure~\ref{fig:halo_occupancy_compare} only shows a single realisation of our model, the results of Figure~\ref{fig:pplf_disk} are marginalised over all uncertainties yet there remains a deficit in the predicted number of faint satellites for the reduced halo occupation model with the fiducial SMHM relation.

This failure of the model motivated us to consider extensions of the SMHM relation for the reduced halo occupation model.
We attempt two extensions which could feasibly permit halos with mass $M_\mathrm{vir}<10^8M_\odot$ to host galaxies which could populate the faintest bin, however find that neither scatter (panel \textit{e} of Figures \ref{fig:pplf_nodisk} and \ref{fig:pplf_disk}) nor a broken power-law SMHM relation (panel \textit{f} of Figures \ref{fig:pplf_nodisk} and \ref{fig:pplf_disk}) are successful in reconciling this model with the data.
The reason for this failure is demonstrated in the bottom-left panel of Figure~\ref{fig:halo_occupancy_compare}, which shows a realisation of our model with the broken power-law SMHM relation.
Since occupation probability decreases with mass in this model, a very small fraction low-mass subhaloes host a luminous galaxy, and those that do are likely to be at large distances -- where subhalos are more numerous -- and hence un-observable.
The same reasoning explains why adding scatter does not work in this case.

We attempt one final extension of this model, where we modify the halo occupancy model from Equation~\ref{eqn:haloocc} such that below $M_\mathrm{vir}<10^8M_\odot$ the halo occupation probability has slope $p_\mathrm{HO} \propto M_\mathrm{vir}^\gamma$, with a prior on this slope uniform between $0<\gamma<2$ and we retain a broken power law SMHM relation.
As demonstrated in the bottom-right panel of Figure~\ref{fig:halo_occupancy_compare}, by allowing the occupation probability to plateau to 0.1 below $M_\mathrm{vir}=10^8M_\odot$ and permitting a break in the SMHM relation, there is a non-negligable probability that $M_\mathrm{vir}<10^8M_\odot$ halos can host faint satellites and that they are nearby enough to be observable.
By modifying halo occupation model in this way, the posterior predictive distribution on the number of satellites in the faintest bin changes from $N<1(6)$ with 68(95\%) credibility (i.e. the prediction shown in panel \textit{f} of Figure~\ref{fig:pplf_disk}) to $N<2(7)$.
This is slightly more commensurate with the observations of $N=5$, however the gain is small.
In any case, we note that this scenario is somewhat contrived since models of reionisation induced suppression of galaxy-formation \citep[e.g.][]{sawala16} typically predict that occupation fraction strictly decreases with decreasing halo mass.

\begin{figure}
\includegraphics[width=\columnwidth]{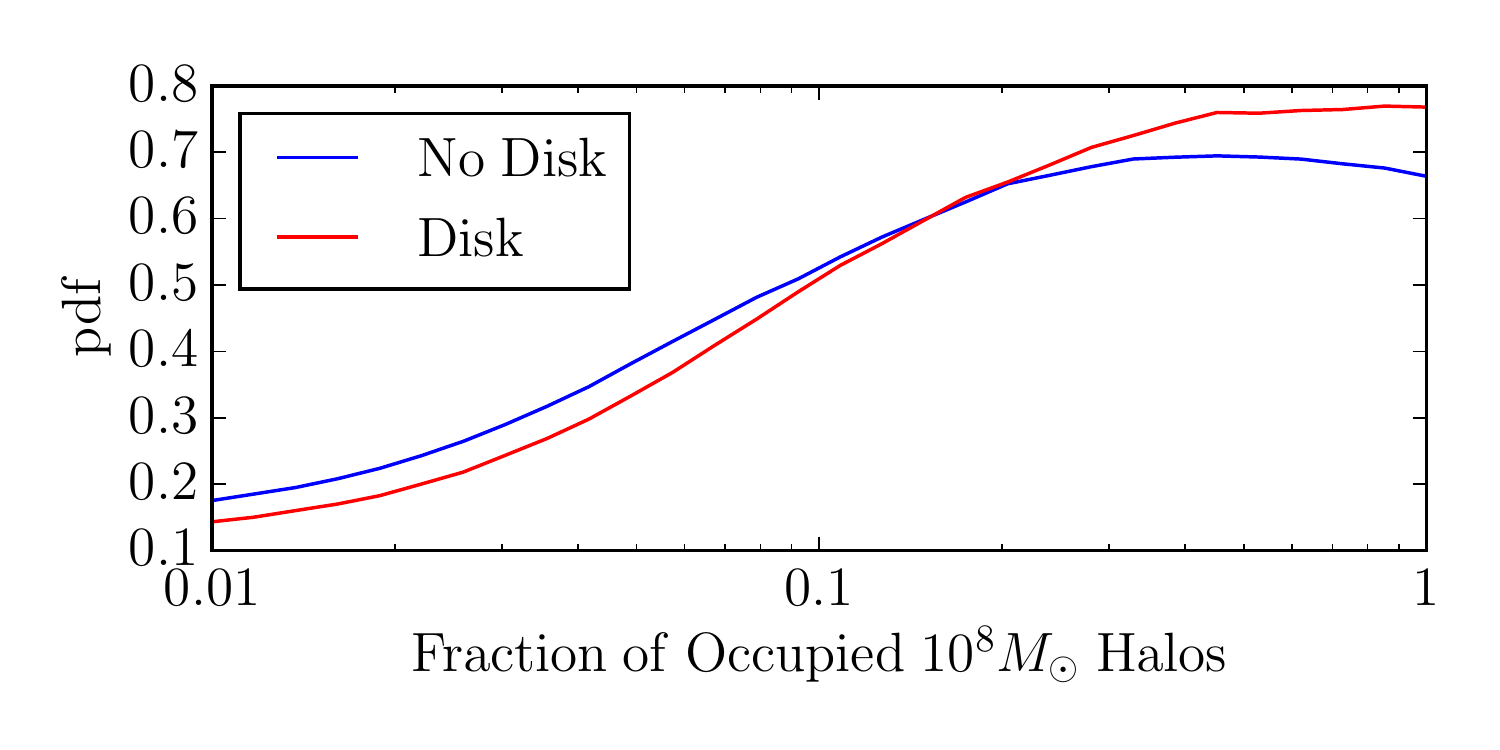}
\caption{
Posterior pdf on occupation fraction for $10^8M_\odot$ halos.
}
\label{fig:frac8}
\end{figure}

With all of these model extensions failing, we conclude that the data somewhat disfavour any scenario where no more than 10\% of $10^8 M_\odot$ halos host a luminous galaxy, signalling a tension with the \citet{sawala16} prediction of how reionisation inhibits galaxy formation in low-mass halos.
To determine what level of halo occupancy our model does in fact require to reproduce the MW population, we widen the range of allowed fraction of occupied $10^8 M_\odot$ halos, $f_8$, to be $0.01<f_8<1$ (prior uniform in log in this range), assigning stellar mass according to our fiducial SMHM relation.
Figure~\ref{fig:frac8} shows posterior pdf on $f_8$.
It is similar with and without a disk, takes a median value $\sim0.2$ and plateaus to a maximum in the range $0.4<f_8<1$.
Thus we conclude that a 40\% occupation fraction can reproduce the data as well as 100\% occupation.

\subsection{Evolution of the Filtering Mass}
\label{ssec:filtering_mass}

\begin{figure}
\includegraphics[width=\columnwidth]{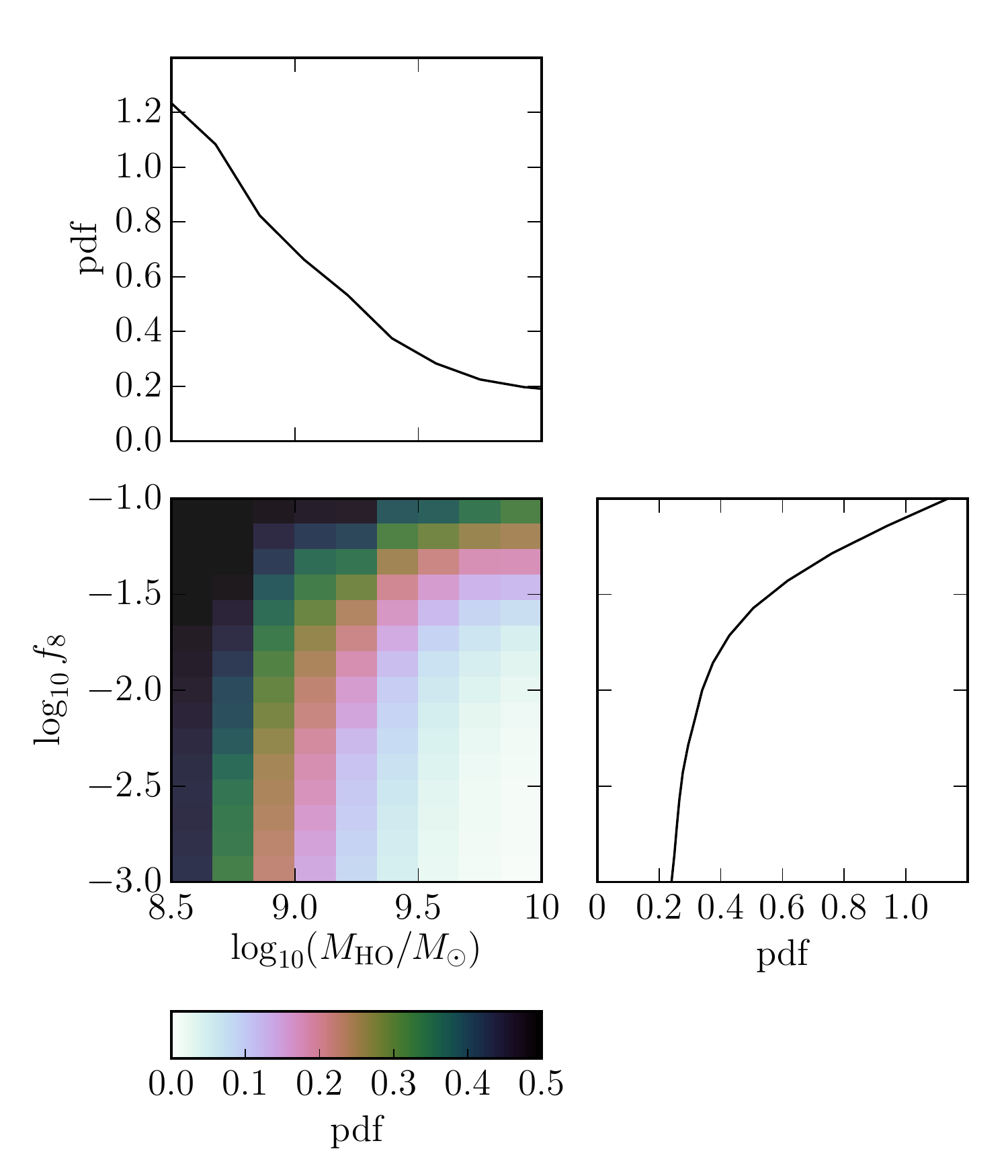}
\caption{
Posterior pdf on the threshold mass below which occupation probability falls below 1 ($M_\mathrm{HO}$) and the fraction of occupied $10^8M_\odot$ halos ($f_8$).
}
\label{fig:filtering_mass}
\end{figure}

Our model of halo occupation is parametrised by the fraction $f_8$ of occupied $10^8M_\odot$ halos, and also a mass $M_\mathrm{HO}$ above/below which occupation probability is equal to/less than one.
Our joint posterior constraints on these parameters are shown in Figure~\ref{fig:filtering_mass}.
As well as constraining $f_8$, as previously discussed, the abundance of MW satellites also constrain the mass threshold $M_\mathrm{HO}$, preferring values $M_\mathrm{HO}<10^9M_\odot$.
If occupation probability falls below one for subhalos more massive than $10^9M_\odot$, our model under-predicts the number of of faint satellites.

The threshold halo mass for efficient galaxy formation is predicted to exhibit strong evolution, e.g. from the cosmological, hydrodynamical simulations of \citet{okamoto08} and \citet{sawala16}.
Though our definition of $M_\mathrm{HO}$ is not directly comparable to the thresholds used in either of these other works, if we nonetheless compare our constraint to the threshold mass predicted in those works, we see that thresholds $M_\mathrm{HO}<10^9M_\odot$ are predicted to occur at redshifts $z>2$.
At later times, even more massive subhalos become liable to not host any luminous galaxy.

\subsection{Predicted luminosity function within 300 kpc}
\label{ssec:predict_300kpc}

Our claim that the the fraction of $10^8M_\odot$ halos which host a luminous galaxy is likely to be above predictions from \citet{sawala16} currently rests upon a small number (5) of galaxies in the faintest bin of our MW satellite luminosity function.
Future detections, or lack thereof, of more and fainter satellite galaxies will be required to draw a firmer conclusion.

\begin{figure*}
\includegraphics[width=\textwidth]{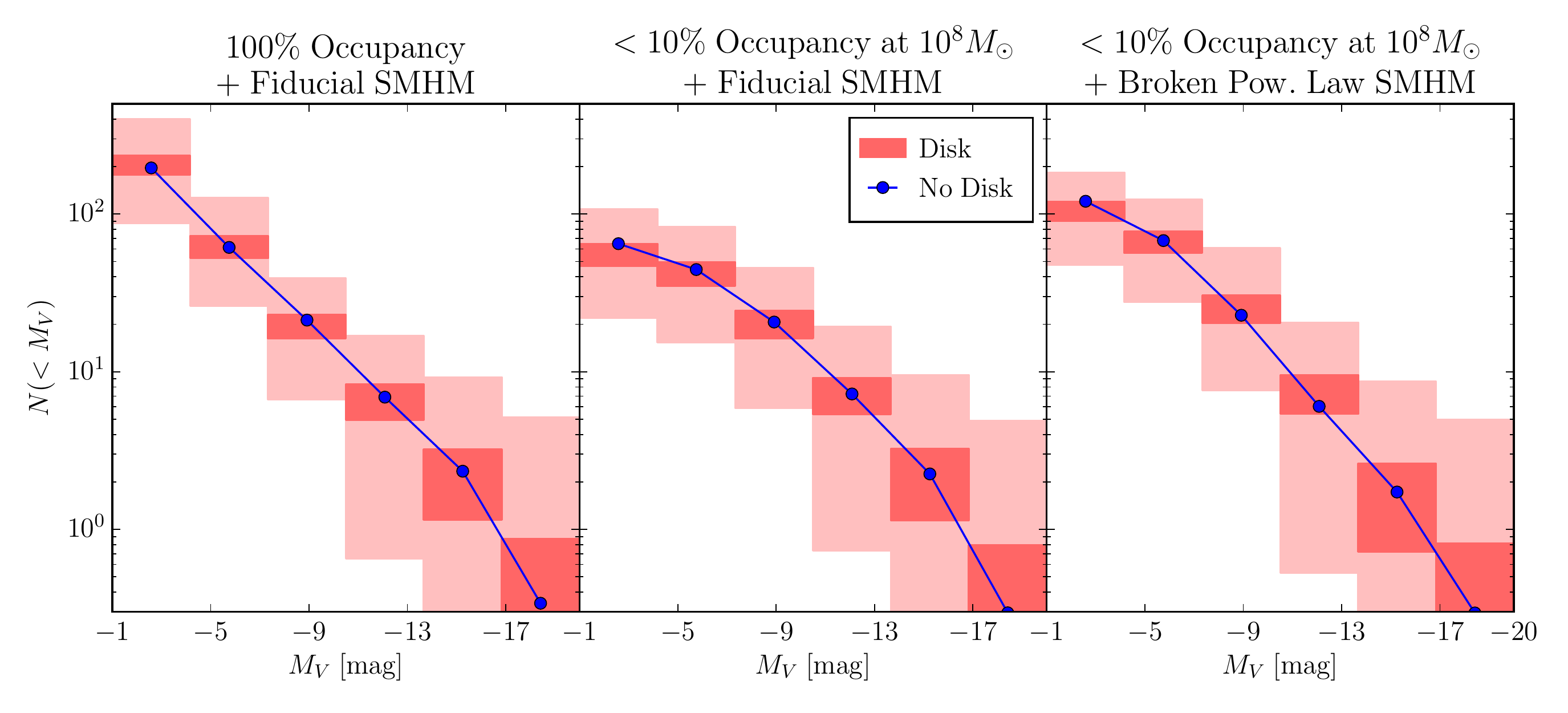}
\caption{
	Predicted luminosity functions (cumulative) of satellite galaxies within 300 kpc.
	Each panels shows a different combination of halo occupation model and SMHM relation: 100\% halo occupation and fiducial SMHM (left), occupation less than $10\%$ for $10^8M_\odot$ halos and the fiducial SMHM (center), and occupation less than $10\%$ for $10^8M_\odot$ halos and the broken-power law SMHM (right).
	Reduced occupation fractions mimic the effects of reionisation as predicted by \citet{sawala16}.
	Each panel shows 1$\sigma$(2$\sigma$) predictions when we include the effect of the disk (dark/light shaded regions), and the median prediction for models with no disk (blue lines).
	}
\label{fig:predicted_lf}
\includegraphics[width=\textwidth]{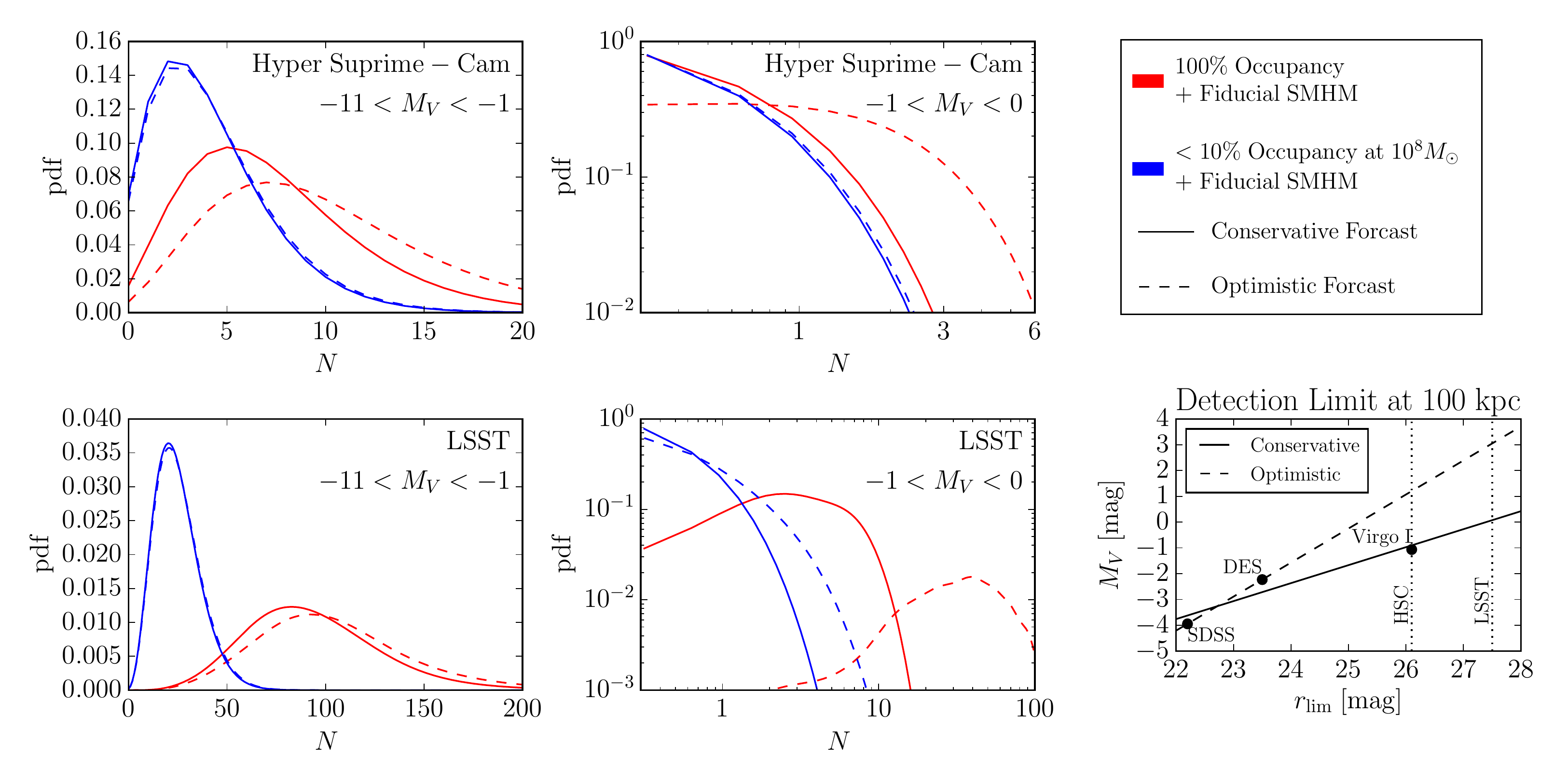}
\caption{
	Predicted number of satellite galaxies detectable in upcoming surveys (within 300 kpc, including known ones).
	The four-panel grid on the left shows pdfs for number of detections: the top row for the 1,400 deg$^2$ Hyper Suprime-Cam survey, the bottom for the LSST assuming full-depth coverage of 20,000 deg$^2$; the left column for galaxies with absolute magnitude in the range $-11<M_V<-1$, the right for $-1<M_V<0$.
	Each panel shows four predictions: red lines assume all subhalos host a luminous galaxy, blue that occupation fraction is no greater than 10\% at $M_\mathrm{vir}=10^8M_\odot$; solid lines assume a conservative forecast of the detection limits, dashed lines an optimistic forecast.
	These forecasts are described in the bottom right panel: it shows $M_V$ for the faintest galaxy detectable at 100 kpc as a function of a survey's $r$-band limiting magnitude.
	The optimistic forecast extrapolates the gain between SDSS and DES to future surveys; the conservative forecast assumes that the one detection so far made in Hyper Suprime-Cam (Virgo I) defines the limit of that survey.
	}
\label{fig:predict_surveys}
\end{figure*}

Figure~\ref{fig:predicted_lf} shows the predicted cumulative luminosity functions for MW satellites within 300 kpc.
The dark/light shaded regions show the 1/2$\sigma$ posterior predictive distributions (calculated using a modified Equation~\ref{eqn:pp_lumfunc}) for models with a disk, and are marginalised over different simulated halos and uncertainties in the SMHM relation.
We show this for three different models.
The left panel shows the prediction assuming 100\% halo occupation and the fiducial SMHM relation.
For this we predict a luminosity function rising continually to $M_V=-1$, and a total number of luminous satellites in the range 178(87)-235(402) for 1$\sigma$(2$\sigma$) credibility interval.
This is in broad agreement with the predictions from both \citet{tollerud08} and \citet{koposov09}, though our quoted uncertainties are larger than these previous works since we include the uncertainty in the SMHM relation.
The central panel shows the predictions when we restrict occupation fraction to 10\% at $M_\mathrm{vir}=10^8M_\odot$ but still for the fiducial SMHM relation.
The predicted luminosity function flattens at the faint end and the predicted total number is between 47(22)-65(107) for 1$\sigma$(2$\sigma$) credibility interval.
For the right panel, we retain the reduced occupation fraction however now use a broken power law SMHM relation, which allows faint satellites to be hosted in less massive subhalos than when we use the fiducial SMHM relation.
This increases the predicted number of faint satellites, however not to the extent that is predicted for 100\% occupation; total predicted numbers are 90(48)-120(183) for 1$\sigma$(2$\sigma$) credibility interval.

In all three panels, we also show the median prediction for the model without a disk (blue lines), and find that the predictions for the disk and no-disk models are very similar.
Given the stark differences in subhalo number density between disk/no-disk simulations shown in Figure~\ref{fig:radial_prof}, this result is somewhat surprising.
The reason for this is discussed in Section~\ref{sec:disk_nodisk}.

\subsection{Predicted number detectable in future surveys}
\label{ssec:predit_survey}

We now predict the number of satellite galaxies which will be detectable in the next generation of deep imaging surveys.
These predictions are obtained by convolving the predicted luminosity function from Section~\ref{ssec:predict_300kpc} with estimates of the footprint and depth of upcoming surveys.
We do this first for the Subaru/Hyper Suprime-Cam survey \citep[HSC,][]{hypersuprimecam}, a presently ongoing survey with a 5$\sigma$ point source $r$-band limiting magnitude of 26.1 mag, covering a $\sim1,400$ deg$^2$ area.
We also make predictions for the Large Synoptic Survey Telescope \citep[LSST,][]{lsst} which is scheduled to begin operations in 2022.
This will have a final, co-added 5$\sigma$ point source $r$-band limiting magnitude of 27.5 mag extending over at least 20,000 deg$^2$.

To estimate the  satellites that will be detectable in future surveys we consider the most extreme detections which have been made to-date.
We first assume that the distance dependence of limiting absolute magnitude for all imaging surveys is equal to that calculated for SDSS (i.e. we assume the slope of Equation~\ref{eqn:selc_func} applies univerally).
Then, given the distance and absolute magnitude of all galaxies discovered in a given survey, we find which one lies closest to the detection limit assuming the above distance-dependence (i.e. given the slope of Equation~\ref{eqn:selc_func}, we find which galaxy requires the most extreme intercept).
Given this slope and intercept, we calculate the faintest $M_V$ that would be observable in a particular survey at 100 kpc.
In the bottom right panel of Figure~\ref{fig:predict_surveys}, we show this for SDSS and DES as a function of the surveys' $r$-band limiting magnitudes.
As an optimistic forecast of the performance of future surveys, we extrapolate this relation to fainter magnitudes (dashed line).

We make a further set of predictions which do not rely on such an extrapolation.
To do this, we assume that the one galaxy already discovered in the first 100 deg$^2$ of data collected by the HSC lies at the detection threshold for that survey.
This galaxy, Virgo I \citep{homma16} has an estimated distance of $\sim87$ kpc, and is one of the faintest known with an absolute magnitude $M_V=-0.8\pm0.9$.
Using the same process described above, we add a further data point to the bottom right panel of Figure~\ref{fig:predict_surveys}, and use a linear least-squares regression line fit to all three data points to give our conservative forcast on future survey performance (solid line).

Using these forecasts, we predict the number of satellite galaxies within 300 kpc which will be detectable in HSC and LSST.
These numbers include already known satellite galaxies.
The four panel grid on the left of figure Figure~\ref{fig:predict_surveys} shows predictions for HSC/LSST (top/bottom row), in two different ranges of absolute magnitude (left/right).
Each panel shows four posterior predictive distributions (calculated using a modified Equation~\ref{eqn:pp_lumfunc}) for the two forecasts on survey performance (solid/dashed lines) and two models of halo occupation (red/blue lines).
For all predictions shown we assume the fiducial SMHM relation.
The uncertainty in the predictions arises from both halo-to-halo scatter and uncertainty in the SMHM relation.

The predictions show that upcoming surveys will be informative in discriminating between different models of halo occupation and hence the physical models of reionisation which underpin them.
In the range $-11<M_V<-1$ (left column) we conservatively (optimistically) predict that HSC should discover $5\pm4$ ($7\pm5$) galaxies assuming 100\% occupancy, compared to $2\pm2$ ($2\pm2$) assuming restricted halo occupancy.
The differences between the models are amplified for LSST, where the conservative (optimistic) predictions are $83\pm32$ ($93\pm36$) for 100\% occupancy and $20\pm11$ ($21\pm11$) for restricted occupancy.
The reason that the optimistic forecasts are not larger than the conservative ones for the model with restricted halo occupation is that the conservative estimate of the limiting magnitude at 300 kpc for both surveys ($M_{V,\mathrm{lim}}^{300\;\mathrm{kpc}}=-3/-2$ mag for HSC/LSST) is inside the regime where the predicted luminosity function for this model begins to flatten (central panel of Figure~\ref{fig:predicted_lf}).
Therefore, though the optimistic forecast of survey performance would allow us to detect $M_V=-1$ galaxies out to distances well beyond current capabilities ($D_\mathrm{lim}^{-1\;\mathrm{mag}}=97/176$ kpc for HSC/LSST) this model simply predicts that no such galaxies should exist.
This is consistent with our claim that restricted halo occupation models predict fewer faint satellites than are observed in SDSS (see Section~\ref{subsubsec:ho_pplf}).

We also predict the number of detections for $-1<M_V<0$ satellites (central column), i.e. probing fainter than the faintest of currently known ultra-faint dwarf galaxies.
These predictions are shown in log-log space since many of them peak very sharply at zero.
This is always the case for the model with restricted halo occupancy (both surveys, both forecasts) which, as discussed above, is a direct consequence of such faint galaxies simply not existing in this model.
They \textit{are} predicted to exist assuming 100\% halo occupancy, and we predict some number should be detectable: $1\pm0.6$ for HSC (optimistic), $3\pm2/39\pm23$ for LSST (optimistic/conservative).

What then, of the one galaxy already discovered in the first 100 deg$^2$ of HSC data, the $M_V=-0.8\pm0.9$ Virgo I galaxy \citep{homma16}?
Taking its central $M_V$ value as gospel, Figure~\ref{fig:predict_surveys} would suggest that discovery this already rules out the restricted halo occupancy model (blue lines, top central panel).
A caveat to this conclusion is that here we only show predictions for the fiducial SMHM model.
Given a more flexible SMHM model, which allows faint galaxies to be hosted in increasingly low-mass halos, it is possible that some $M_V>-1$ satellites exist even if we assume restricted halo occupation.
This can be seen in Figure~\ref{fig:predicted_lf}, where the predicted number of satellite galaxies assuming a restricted halo occupation model and a broken power law SMHM relation (right panel) is greater than the number predicted assuming the fiducial SMHM relation (center panel); it is still less, however, than the prediction assuming 100\% halo occupation (left panel).
Similarly for Figure~\ref{fig:predict_surveys}, the predictions for 100\% occupancy (red) and $<10\%$ occupancy for the fiducial SMHM relation (blue) bracket the predicted observable numbers assuming $<10\%$ occupancy and a broken power law SMHM relation (not shown).
Given this, we believe that if two or more $M_V>-1$ galaxies are additionally discovered in HSC (i.e. a total $N\geq3$), this would be difficult to accommodate in models where $<10\%$ of $10^8M_\odot$ halos host a luminous galaxy.

\section{The SMHM Relation}
\label{sec:SMHM}

We now ask what constraints we can place on the connection between galaxy stellar mass and halo mass at these low-mass scales.
To this end, we calculate the posterior predictive distributions on the SMHM relations.

\subsection{Inferred Stellar Mass at Fixed Halo Mass}

\begin{figure*}
\includegraphics[width=\textwidth]{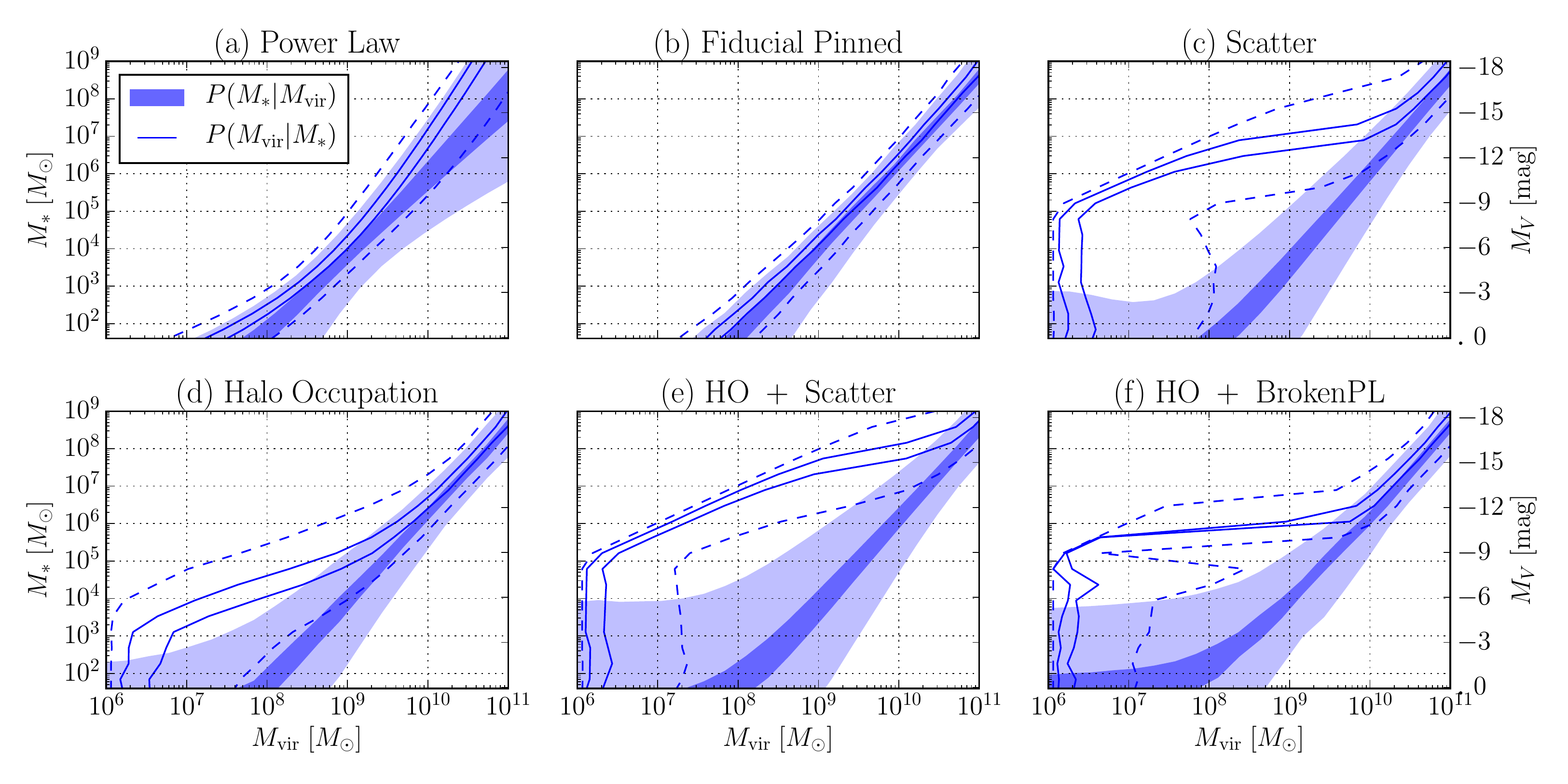}
\caption{
Posterior predictive SMHM relations for simulations with no disk.
These are shown for six different SMHM models as indicated by panel titles.
The dark/light shaded regions show the 68/95\% confidence intervals for stellar mass at a fixed halo mass, to be read off vertically.
The solid/dashed lines show the 68/95\% confidence intervals for halo mass at a fixed stellar mass, to be read off horizontally.
}
\label{fig:SMHM_nodisk}
\includegraphics[width=\textwidth]{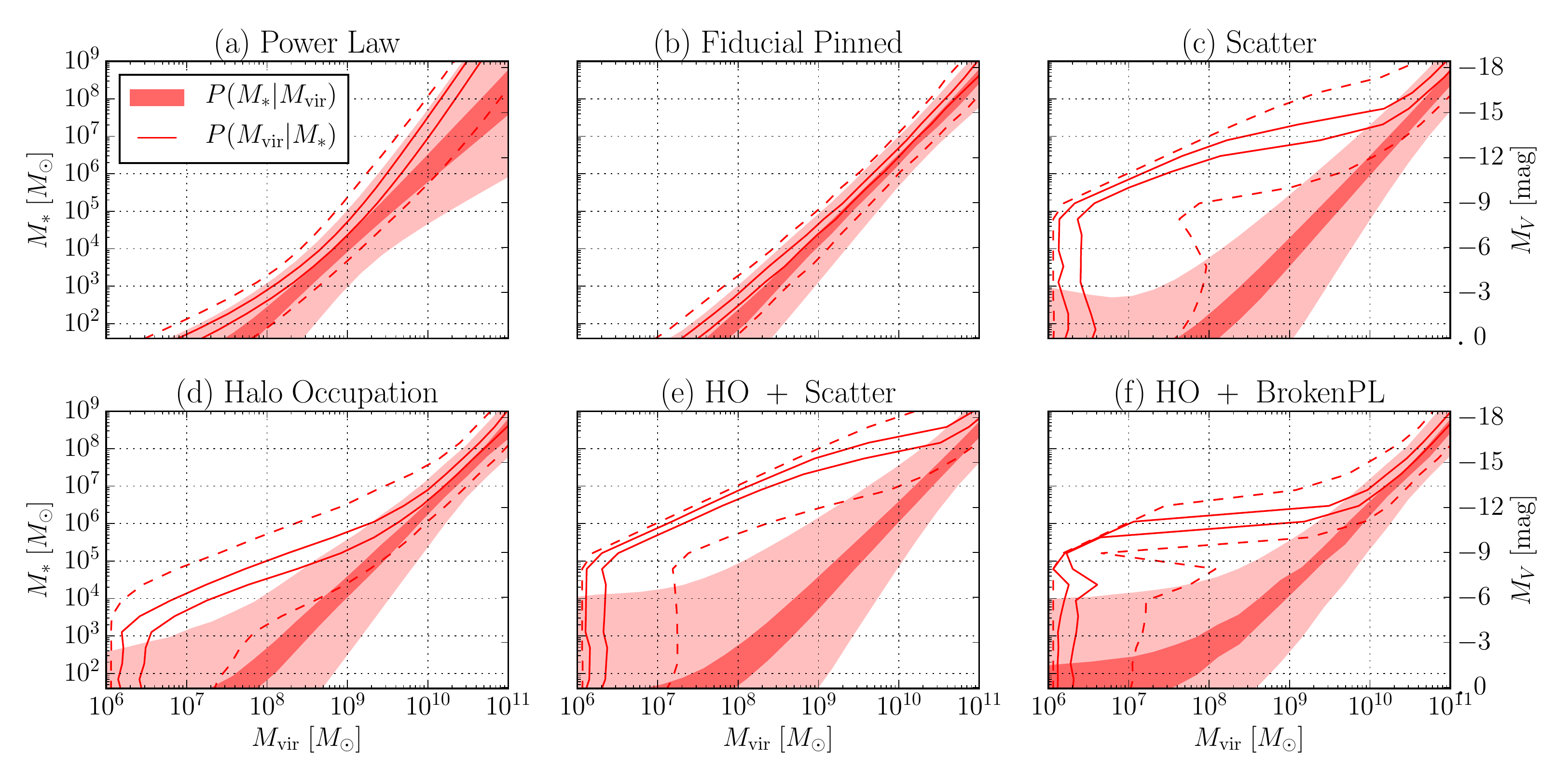}
\caption{As Figure \ref{fig:SMHM_nodisk} but for simulations with a disk.}
\label{fig:SMHM_disk}
\end{figure*}

For a model defined by its SMHM relation $P_\Theta(M_*|M_\mathrm{vir})$ (which is a pdf on stellar mass as a function of halo mass), the posterior predictive distribution on the stellar mass for fixed halo mass $M_\mathrm{vir}$, conditioned on observed data $\mathbf{X}$ is given by
\begin{equation}
P(M_*|M_\mathrm{vir},\mathbf{X}) = \int P_\Theta(M_*|M_\mathrm{vir}) P(\Theta|\mathbf{X}) \; \mathrm{d}\Theta.
\label{eqn:pp_MstarMvir}
\end{equation}
Figures \ref{fig:SMHM_nodisk} and \ref{fig:SMHM_disk} show this quantity for models without and with a disk, respectively.
The light/dark filled, colored bands show the 68/95\% credibility intervals (the solid/dashed lines are discussed in Section~\ref{ssec:invert_constraints}.)
Here we discuss the results of using our different SMHM prescriptions.
We defer discussion of the differences between disk/no-disk models to Section~\ref{sec:disk_nodisk}.

\subsubsection{Power Law \& Fiducial models: agree at $M_\mathrm{vir}=10^{11}M_\odot$}

The difference between our basic power law SMHM model and our fiducial model is that, for the latter, we use an informative prior on the stellar mass of a $10^{11}M_\odot$ halo based on abundance matching studies at higher mass scales.
As expected, introducing an informative prior tightens our constraints (compare panels \textit{a} and \textit{b} of Figure \ref{fig:SMHM_nodisk} or \ref{fig:SMHM_disk}).
More interestingly, we see that the constraints on $P(M_*|M_\mathrm{vir}=10^{11}M_\odot)$ from both of these models are consistent with one another at the $1\sigma$ level.
This means that the SMHM relation as inferred from just the 25 satellite galaxies in Table~\ref{tab:data} matches on to the SMHM relation inferred from samples of thousands of galaxies from \citet{behroozi13} and \citet{moster13}.
Note that all other models discussed use the informative prior on $P(M_*|M_\mathrm{vir}=10^{11}M_\odot)$.

\subsubsection{$P(M_*|M_\mathrm{vir})$ in Scatter Models}
\label{subsubsec:scatter_smhm}

The $1\sigma$ constraints on $P(M_*|M_\mathrm{vir})$ are steepest for the scatter model (shaded regions, panel \textit{c} of Figure \ref{fig:SMHM_nodisk} or \ref{fig:SMHM_disk}).
This corroborates the result from \citet{garrison17} that including scatter in the SMHM relation leads to a steepening of the inferred relation between mean stellar mass for a given halo mass.
This is a result of an Eddington bias: due to the rising subhalo mass function in $\Lambda$CDM, more low mass subhalos will be up-scattered to a given stellar mass than high mass subhalos down-scattered.
Therefore, a steep mean SMHM relation with scatter can produce an observed luminosity function similar to a shallow SMHM relation with no scatter.

\subsubsection{$P(M_*|M_\mathrm{vir})$ in Halo Occupation Models}
\label{subsubsec:ho_smhm}

A qualitatively different trend is observed for the halo occupation models (shaded regions, panels \textit{d,e} and \textit{f} of Figure \ref{fig:SMHM_nodisk} or \ref{fig:SMHM_disk}).
The constraints on $P(M_*|M_\mathrm{vir})$ for these models have broad wings extending to higher stellar masses for a given halo mass.
This is especially clear looking at the region $P(M_*<10^4M_\odot|M_\mathrm{vir}<10^8M_\odot)$ which contains scant probability for models with 100\% occupancy, but becomes increasingly more likely for models where we limit the occupation fraction of $10^8M_\odot$ subhalos to 10\%.
This happens because, having imposed low occupation fractions, the model tries to place galaxies in increasingly less massive hosts.
We also see the strength of this effect increase going from panel \textit{d}, to \textit{e}, to \textit{f}.
This is because introducing extra flexibility in the SMHM models (i.e. scatter in panel \textit{e} and a broken power law in panel \textit{f}) allows the model extra freedom to access increasingly less massive subhalos.

\subsection{Inferred Halo Mass at Fixed Stellar Mass}
\label{ssec:invert_constraints}

All of the above discussion has concerned the our inference on stellar mass for a fixed halo mass, i.e. $P(M_*|M_\mathrm{vir})$.
Given that we observe stellar content, however, the quantity we are more interested in is $P(M_\mathrm{vir}|M_*)$, i.e. our inference on halo mass for a fixed stellar mass.
Using Bayes' theorem, these are related by
\begin{equation}
P(M_\mathrm{vir}|M_*) \propto P(M_*|M_\mathrm{vir}) P(M_\mathrm{vir}),
\label{eqn:pp_MvirMstar}
\end{equation}
where $P(M_\mathrm{vir})$ is the prior on the subhalo mass.
Assuming no additional knowledge about which halo hosts a given galaxy, this prior is given by the halo mass function.
This non-trivial distinction between $P(M_*|M_\mathrm{vir})$ and $P(M_\mathrm{vir}|M_*)$ is pointed out in footnote 5 of \citet{behroozi13}, and discussed more thoroughly in \citet{dooley16}.
Here, we explicitly calculate the posterior predictive distributions for $P(M_\mathrm{vir}|M_*)$ for each of our SMHM models.
This is given by inserting a factor of $P(M_\mathrm{vir})$ -- where we assume a $\Lambda$CDM mass function -- into the integral in Equation~\ref{eqn:pp_MstarMvir}.

The delineated regions in Figures~\ref{fig:SMHM_nodisk} and ~\ref{fig:SMHM_disk} show our inference on $P(M_\mathrm{vir}|M_*)$ (solid/dashed lines show 68/95\% credible regions).
For the power law and fiducial models (panels \textit{a} and \textit{b}), we see that $P(M_\mathrm{vir}|M_*)$ is largely coincident with $P(M_*|M_\mathrm{vir})$.
This is not true for the scatter model (panel \textit{c}) or halo occupation models (panel \textit{d,e} and \textit{f}).
Although the constraints on $P(M_*|M_\mathrm{vir})$ for these models are qualitatively similar to those found for the fiducial, the inferred halo mass for a given stellar mass is pushed to low values.
This difference is due to the stochastic way halos are assigned stellar mass when we include scatter and unoccupied subhalos, which results in low-mass halos ($M_\mathrm{vir}\sim10^7M_\odot$) having a non-zero (but small) probability of hosting bright galaxies ($M_*\sim10^6M_\odot$).
Given the steeply rising $\Lambda$CDM mass function however, though a given low-mass halo only has a small probability of hosting a bright galaxy, the fact that low-mass halos are far more numerous makes them the most likely host for any given observed bright galaxy.

\subsection{Connection to $z=0$ mass}
\label{ssec:zeq0}

\begin{figure*}
	\centering
	\subfigure{\includegraphics[width=\columnwidth]{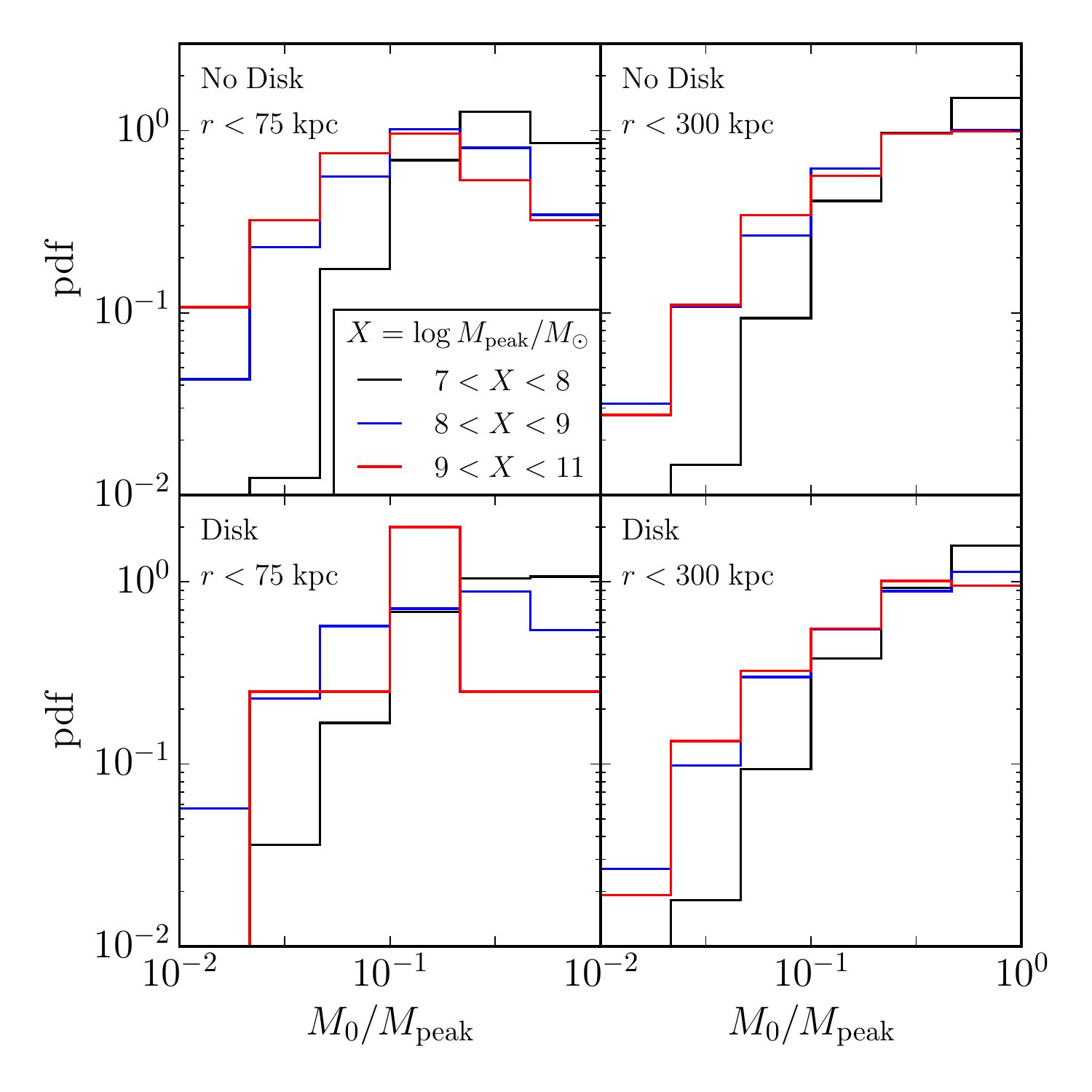}}\quad
	\subfigure{\includegraphics[width=\columnwidth]{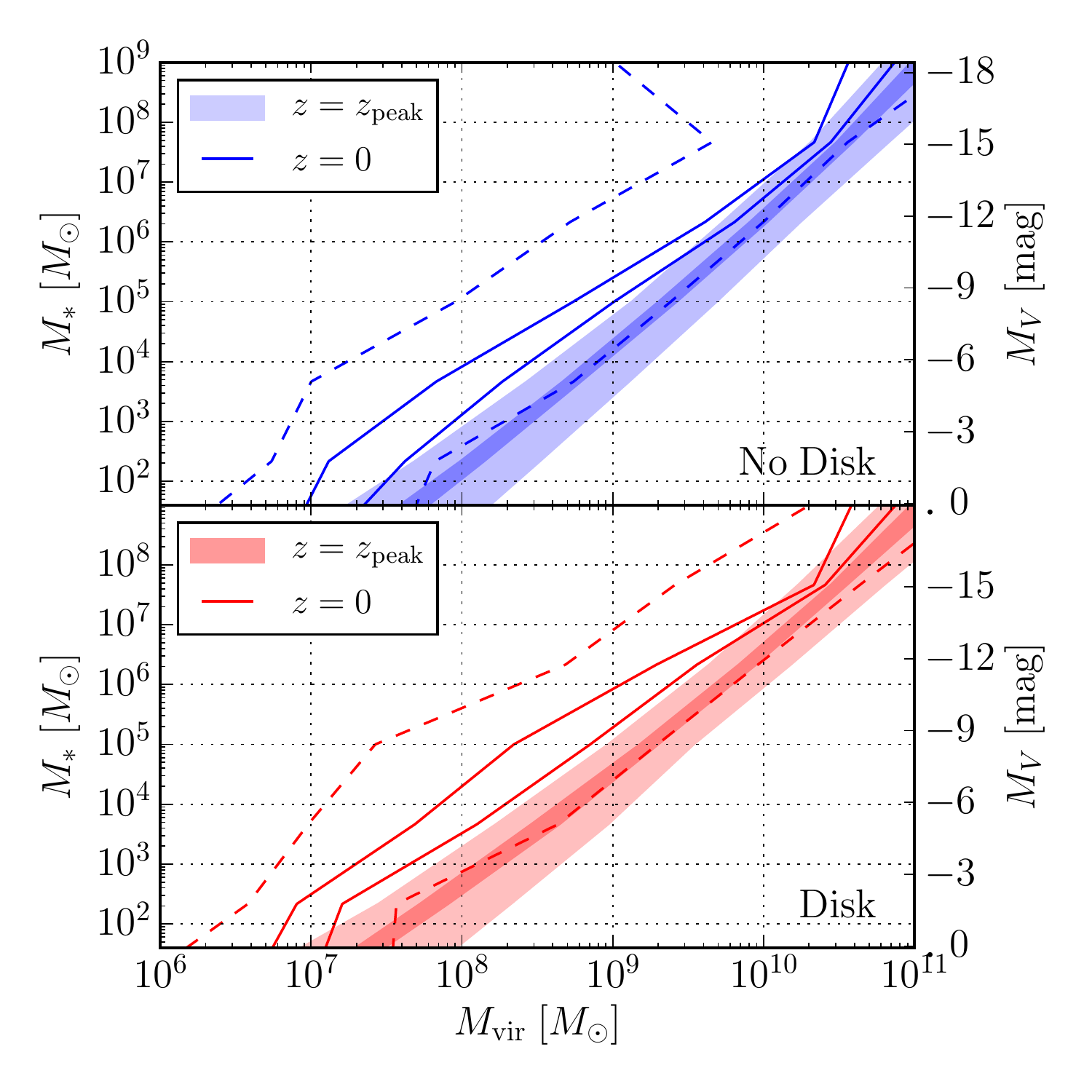}}
	\caption{
	Transforming constraints from $z_\mathrm{peak}$ to $z=0$.
	The left panel shows histograms of the fractional mass-loss undergone by subhalos within 75/300 kpc of their host (left/right columns) for models without/with a disk (top/bottom rows) in three different mass bins (see legend).
	In the right panel, we show 1$\sigma$/2$\sigma$ constraints on $P(M_*|M_\mathrm{vir})$ at $z=z_\mathrm{peak}$ (dark/light colored bands) and the constraints transformed to $z=0$ (solid/dashed delineated regions) for models without/with a disk (top/bottom rows).
	}
\label{fig:zeq0}
\end{figure*}

The halo property which we have used as a proxy for galaxy stellar mass is the peak virial mass.
We would also like to predict the current halo mass of satellite galaxies after accounting for the effects of tidal stripping from the host halo, since the $z=0$ value is required for predicting the dark-matter annihilation signal (though such a signal would also be sensitive to the internal mass profile).

In order to transform our constraints to $z=0$, we quantify the mass-loss function of subhalos in our simulations.
We calculate the pdf on the $z=0$ mass of a subhalo of a given peak mass, subject to the constraint that the $z=0$ halo-centric distance of the subhalo is $r<R$ for some value $R$, i.e. $P(M_\mathrm{vir}^0|M_\mathrm{vir}^\mathrm{peak}, r<R)$.
The left panel of Figure~\ref{fig:zeq0} shows mass-loss functions averaged over all of our $N$-body simulations: we can see that mass-loss is (i) stronger for subhalos which are closer to the host (compare left/right columns), (ii) greater, in fractional terms, for halos with larger peak masses (compare black/blue/red distributions), and (iii) largely independent of the presence of the disk for the population of surviving subhalos (compare top/bottom rows).
The posterior predictive distribution on $z=0$ halo mass for fixed stellar mass $M_*$ conditioned on observed data $\mathbf{X}$ is then given by,
\begin{align}
P(M_\mathrm{vir}^0|M_*,\mathbf{X}) \propto \int  P(M_\mathrm{vir}^0 & |M_\mathrm{vir}^\mathrm{peak}, r<R(M_*)) \\	\nonumber
	& P(M_\mathrm{vir}^\mathrm{peak}|M_*,\mathbf{X}) \;\; \mathrm{d}M_\mathrm{vir}^\mathrm{peak},
\label{eqn:zeq0}
\end{align}
where the first term in the integral is the mass-loss function, $R(M_*)$ corresponds to the limiting distance for detection of an $M_*$ galaxy (Equation~\ref{eqn:selc_func}), and the second term is the posterior predictive distribution on peak mass (Equation~\ref{eqn:pp_MvirMstar}).

The right panel of Figure~\ref{fig:zeq0} shows the transformation from peak mass constraints (colored bands) to $z=0$ constraints (solid/dashes lines) for our fiducial SMHM model.
To calculate these we tabulate mass-loss functions for six bins of stellar mass in the range $10<M_*/M_\odot<10^9$.
We see that the $z=0$ constraints differ most from the peak mass constraints at the low mass end.
This is because, although low-mass halos undergo fractionally less mass-loss than high-mass halos, faint satellites are only visible nearby and hence are more prone to tidal stripping.
The differential change between peak and $z=0$ constraints is similar for simulations with and without a disk (top/bottom panels).
We note that the $z=0$ constraints are slightly noisy since they have been calculated from mass-loss functions based on only a few subhalos in high- and low-mass bins.

\subsection{The Least Massive Halo}
\label{ssec:leastmassive}

In Figure~\ref{fig:mass_lightest} we show the inference on the mass of the faintest satellite in our sample, i.e. the $M_V=-1.5$ dwarf galaxy Segue I.
The top panel shows our inference on the peak mass, which we convolve with halo mass loss function as described in Section~\ref{ssec:zeq0} to get the inference on $z=0$ mass shown in the bottom panel.
We show this for three SMHM models -- fiducial, scatter and halo occupation -- both with (red) and without (blue) a baryonic disk.
Since we are calculating $P(M_\mathrm{vir}|M_*)$ at a fixed stellar mass, it is necessary to introduce a prior probability distribution on the value of $M_\mathrm{vir}$ (see Equation~\ref{eqn:pp_MvirMstar}).
The left and right halves of this plot shows the results when we assume two different such priors.

The top right quadrant of the plot shows the inference on peak mass assuming a prior given by the $\Lambda$CDM subhalo mass function \citep{springel08}.
These are identical to horizontal slices of the constraints shown by the solid/dashed lines in Figures~\ref{fig:SMHM_nodisk} and ~\ref{fig:SMHM_disk}.
As discussed in Section~\ref{ssec:invert_constraints}, for models which include stochasticity in the SMHM relation -- i.e. scatter and halo occupantion models -- our inference on the halo mass is pushed to low values since low mass halos are simply more numerous.
Taking a more agnostic approach to the existence of low mass halos, in the top left quadrant of Figure~\ref{fig:mass_lightest} we show $P(M_\mathrm{vir}|M_V=-1.5)$ assuming a flat halo mass function.
The model with scatter results in the highest mass constraint (as discussed in Section~\ref{subsubsec:scatter_smhm}) while the effect of the baryonic disk is to reduce the inferred masses by $\sim0.2$ dex.
We take this combination of models (SMHM with scatter, including the disk), under the assumption of a flat halo mass function prior, to give our conservative upper bound on the peak halo mass of the faintest MW dwarf: $M_\mathrm{vir} < 2.4 (13.9) \times10^8M_\odot$ at $1\sigma(2\sigma)$ credible interval.

The bottom half of the plot shows the constraints transformed to $z=0$.
We note that halo masses below $3\times10^7M_\odot$ are not converged in our simulations, so any constraints below this should be treated with caution.
For our main reported result, we again give the most conservative upper bound amongst models with a disk, i.e. the scatter model assuming a flat mass function: $M_\mathrm{vir} < 0.6 (2.1) \times10^8M_\odot$ at $1\sigma(2\sigma)$ credible interval.

\section{The Effect of the MW disk}
\label{sec:disk_nodisk}

The MW disk destroys subhalos in the inner region of the MW halo.
This is clearly shown in Figure~\ref{fig:radial_prof}, where the subhalos number density in the disk models is depleted by a factor $\sim3$ at $0.1r_\mathrm{vir}$ compared to the no-disk models.
This results in a much flatter radial subhalo number density profile, and is consistent with previous works looking into this effect \citep{donghia_et_al_2010}.

Given the destructive power of the disk, it is surprising that most of our results seem to be largely insensitive to whether we use the disk or no-disk simulations.
For example, in Figure~\ref{fig:predicted_lf} we see that the predicted luminosity function for the disk (red bands) and no-disk (blue line) models look very similar.
The reason for this insensitivity is that we require our models to reproduce the observed satellite galaxy luminosity function.
To satisfy this requirement demands that the SMHM relation for disk models is shallower than than the equivalent no-disk model, allowing less massive and more numerous subhalos to act as galaxy hosts in the disk models, compensating for subhalo destruction.

\begin{figure}
\includegraphics[width=\columnwidth]{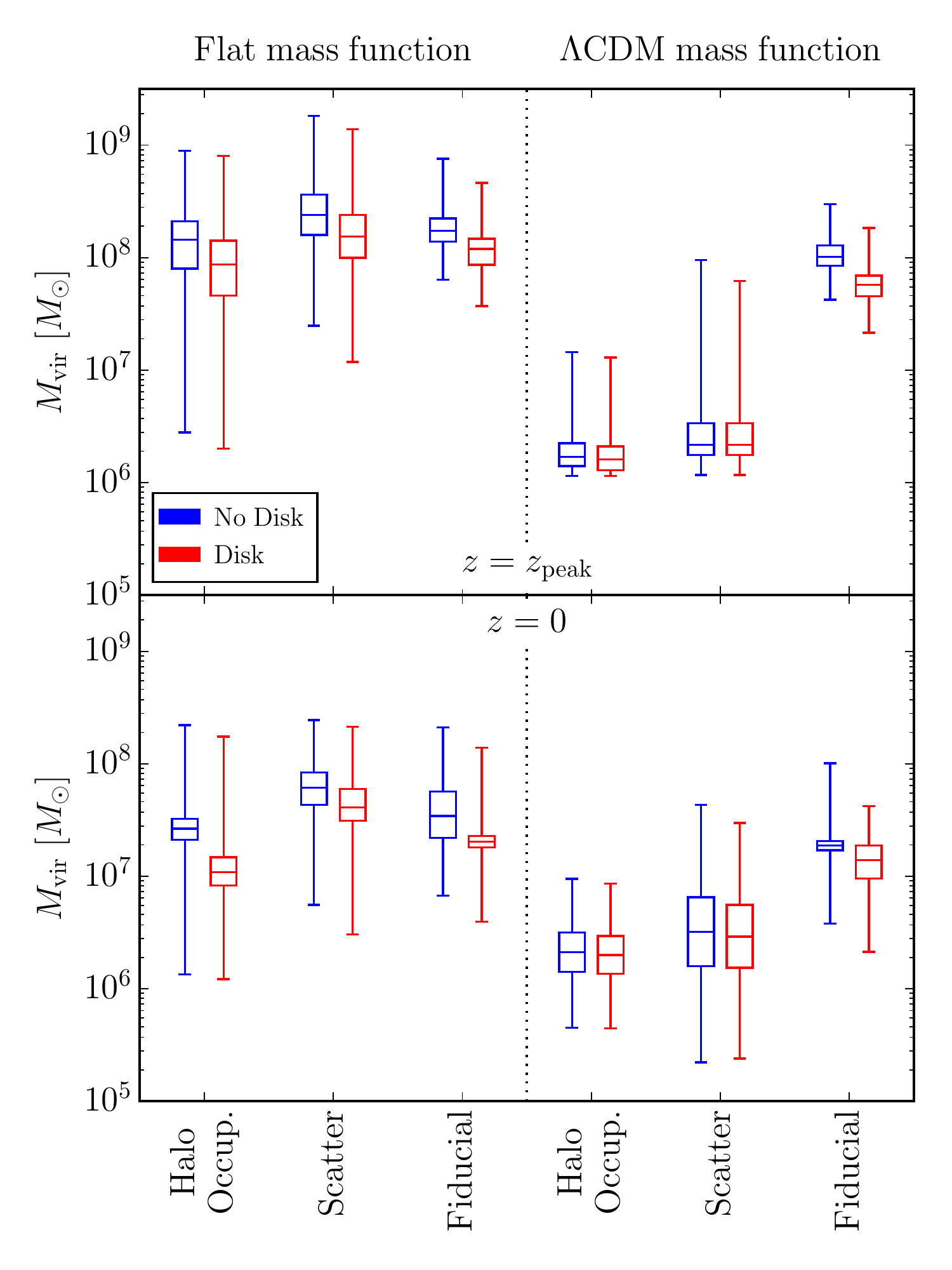}
\caption{
Inference on the halo mass of a $M_V=-1.5$ satellite.
Top half shows inference on peak mass, bottom half $z=0$ mass; for the right half we assume a prior on $M_\mathrm{vir}$ given by the $\Lambda$CDM subhalo mass function, the left half a flat mass function.
Each box-and-whisker plot show the 1- and $2\sigma$ constraints.
We show this for three SMHM models as labeled on the \textit{x}-axis and for models with (red) and without (blue) a disk.
}
\label{fig:mass_lightest}
\end{figure}

The shallower inferred SMHM relation for disk models can be seen by comparing $P(M_*|M_\mathrm{vir})$ in like-for-like panels between Figures~\ref{fig:SMHM_nodisk} and~\ref{fig:SMHM_disk}.
The difference in slopes is small, however it can be seen more clearly in the Figure~\ref{fig:mass_lightest}, which shows the inferred halo mass of the faintest MW satellite galaxy.
For every comparison between disk/no-disk (red/blue) constraints, the disk result is always less massive than that with no-disk, attesting to the shallower SMHM relation.
Though the difference is small (indeed they are $1\sigma$ consistent in all but one of the comparisons), this slight difference in slope is sufficient to compensate for the subhalo depletion by the disk.

\section{Constraints on Warm Dark Matter}
\label{sec:wdm}

Measurements of the matter power spectrum on small scales can discriminate between competing models of dark matter.
In a cold dark matter universe, the mass function of bound dark matter halos continues to rise to arbitrarily small masses \citep{springel08}.
Contrast this to warm dark matter (WMD) cosmologies, in which the dark matter particle remains relativistic for some time after decoupling in the early Universe, leading to a suppression in small scale structure growth at early times \citep[e.g.][]{bode01,avilareese01}.
As discussed in Section~\ref{ssec:leastmassive}, we have placed a 68\% credibility upper bound of $2.4 \times10^8M_\odot$ on the halo mass of the least luminous satellite in our sample.
We now ask what constraints this allow us to place on the nature of the dark matter particle.

\subsection{Free Streaming Approximation}

A simple argument which connects our result to the dark matter particle would assert that our inferred halo mass must lie above the free streaming mass scale of the dark matter particle, since below this initial density perturbations are completely erased.
An effective free streaming mass scale \citep{schneider12} is given by
\begin{equation}
M_\mathrm{fs}=\frac{4\pi}{3}\rho_\mathrm{crit}\left(\frac{\lambda_\mathrm{fs}}{2}\right)^3,
\label{eqn:free_stream_mass}
\end{equation}
where the free streaming length scale $\lambda_\mathrm{fs}$ is given by,
\begin{equation}
\lambda_\mathrm{fs} = 0.049 \left(\frac{m_\mathrm{WDM}}{\mathrm{keV}}\right)^{-1.11} \left(\frac{\Omega_\mathrm{0}}{0.25}\right)^{0.11} \left(\frac{h}{0.7}\right)^{1.22} \; \mathrm{Mpc} \; h^{-1},
\label{eqn:free_stream_length}
\end{equation}
$m_\mathrm{WDM}$ is the warm dark matter particle mass, $\Omega_\mathrm{0}$ is the matter density of the Universe at redshift 0, and $h$ is the Hubble parameter.
Assuming cosmological parameters from Section~\ref{sec:sims} we convert our upper bound on the halo mass into a lower bound on the dark matter particle mass: $m_\mathrm{WDM} > 0.3(0.5)$ keV at $1\sigma(2\sigma)$ confidence.

\subsection{WDM mass function}

The above calculation ignores the fact that WDM leads to significant suppression in power on scales larger than the free streaming length scale.
\citet{schneider12} provide a parametric form of the $\Lambda$WDM mass function in terms of that from $\Lambda$CDM,
\begin{equation}
\left. \frac{\mathrm{d}N}{\mathrm{d}M}\right|_\mathrm{WDM} =
\left. \frac{\mathrm{d}N}{\mathrm{d}M}\right|_\mathrm{CDM}
\left(1 + \frac{M_\mathrm{hm}}{M}\right)^{-\beta},
\label{eqn:wdm_mass_func}
\end{equation}
where the index $\beta$ is fit to simulations, the \textit{half mode mass}
\begin{equation}
M_\mathrm{hm}=\frac{4\pi}{3}\rho_\mathrm{crit}\left(\frac{\lambda_\mathrm{hm}}{2}\right)^3,
\end{equation}
is the mass at which the amplitude of the WDM transfer function reduced by 1/2, and $\lambda_\mathrm{hm} = 13.93 \lambda_\mathrm{fs}$ is the \textit{half mode length}.
We incorporate this into our model in a similar way to our implementation of variable halo occupation.
For each CDM halo we assign a probability that it would exist in a WDM universe according to Equation~(\ref{eqn:wdm_mass_func}).
We then Monte Carlo sample realisations of the subhalo population in a WDM universe, then assign stellar mass using one of the SMHM models described in Section~\ref{ssec:mstarmods}.
Finally, we simultaneously constrain $m_\mathrm{WDM}$ and the parameters of the SMHM model as described in Section~\ref{ssec:probmod}.

\begin{figure}
\includegraphics[width=\columnwidth]{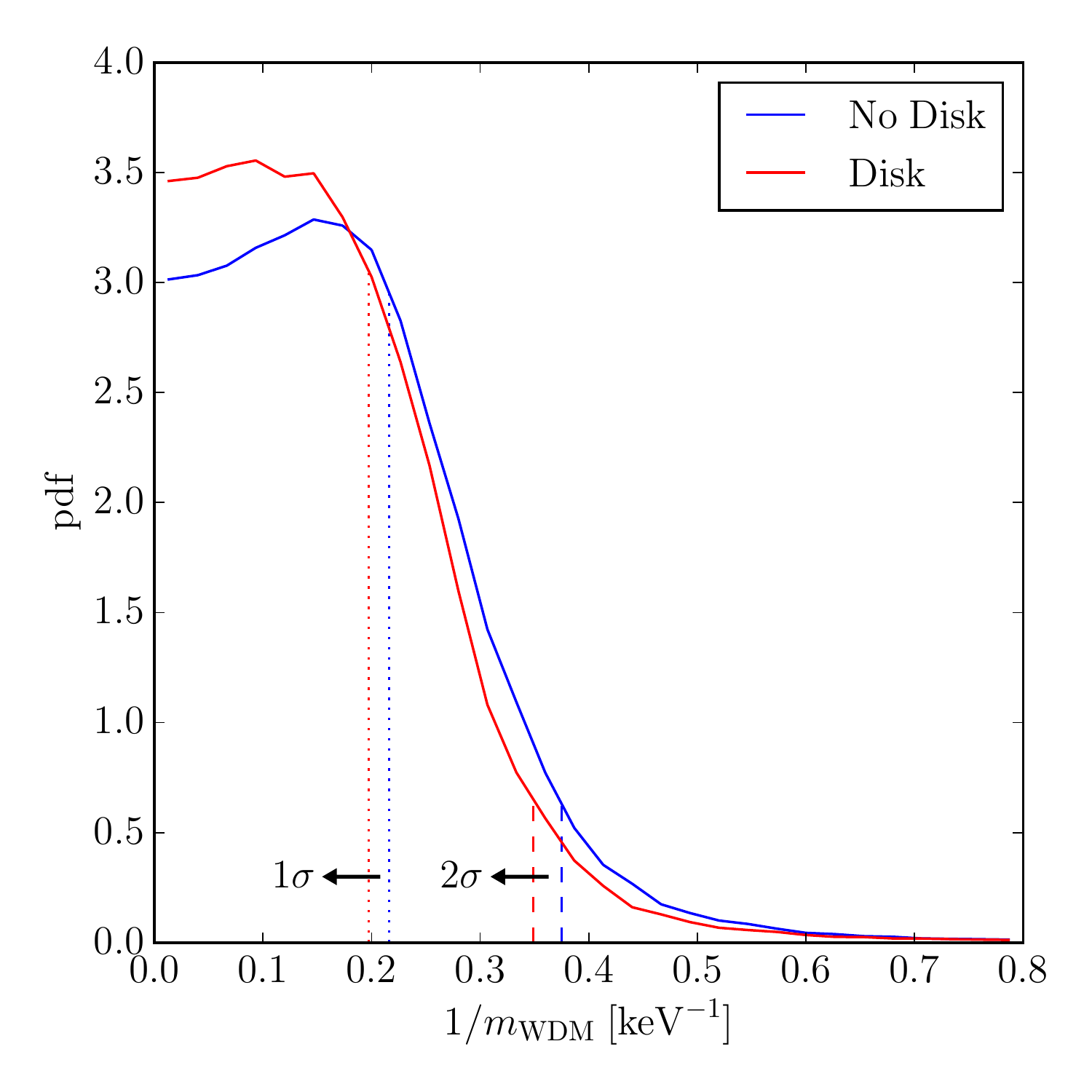}
\caption{
Posterior pdf on one over the mass of a thermal relic dark matter particle for simulations with (red) and without (blue) a disk.
Vertical dotted/dashed lines show our 1$\sigma$/2$\sigma$ upper bounds on $1/m_\mathrm{WDM}$, i.e. lower bounds on $m_\mathrm{WDM}$.
}
\label{fig:mWDM}
\end{figure}

Figure~\ref{fig:mWDM} shows our constraints on $m_\mathrm{WDM}$, assuming a prior on the particle mass uniform in $1/m_\mathrm{WDM}$.
This prior was chosen in order to able to directly compare our results with constraints derived from the Lyman $\alpha$ forest \citep[][though we will discuss the validity of this prior in Section~\ref{sssec:wdm_prior}]{viel08,viel13,baur16}.
It shows the posterior pdf on $1/m_\mathrm{WDM}$ for the simulations with a disk (red) and without (blue), marginalised over the choice of SMHM model (either scatter or fiducial) and the WDM suppression index $\beta$ (either $\beta=1.16$ from \citealt{schneider12} and $\beta=1.3$ from \citealt{lovell14}).
From the simulations with no disk we derive constraints $m_\mathrm{WDM}>4.63(2.67)$ keV at 1$\sigma$(2$\sigma$).
Including subhalo depletion due to the disk, this rises to $m_\mathrm{WDM}>5.07(2.87)$ keV at 1$\sigma$(2$\sigma$).
We discuss these constraints in the context of other works in Section~\ref{sec:discuss}.

\begin{table}
	\centering
	\caption{
	Lower bounds on $m_\mathrm{WDM}$ for different model assumptions.
	In the top half of the table, columns show whether the model contains a disk, the value of $\beta$ from Equation~(\ref{eqn:wdm_mass_func}) controlling the shape of the WDM mass function, the SMHM relation used and the 1/2$\sigma$ lower bounds.
	The bottom half of the table shows how the results vary with MW mass.
	}
	\label{tab:WDM}
	\begin{tabular}{lllcc}
	\multirow{2}{*}{Disk?} 	&	\multirow{2}{*}{$\beta$}	& \multirow{2}{*}{SMHM model} 		& \multicolumn{2}{l}{$m_\mathrm{WDM}/\mathrm{keV}>$} \\
							&								&									& 1$\sigma$	& 2$\sigma$ \\
	\hline
	N      					&   1.16    					& Fiducial   						&        4.49     &  2.62 			\\ 
	N      					&   1.30     					& Fiducial 	  						&        4.69     &  2.74 			\\ 
	N      					&   1.16    					& Scatter           				&        4.57     &  2.59 			\\ 
	N      					&   1.30     					& Scatter           				&        4.75     &  2.71 			\\
	\textbf{N} 				& 	\multicolumn{2}{c}{\textbf{--- Marginalised ---}}				& \textbf{4.63}	  &  \textbf{2.67} 	\\
	Y     					&   1.16    					& Fiducial   						&        4.92     &  2.81 			\\ 
	Y     					&   1.30     					& Fiducial   						&        5.12     &  2.92 			\\ 
	Y     					&   1.16    					& Scatter           				&        5.01     &  2.81 			\\ 
	Y     					&   1.30     					& Scatter           				&        5.21     &  2.93 			\\
	\textbf{Y} 				& 	\multicolumn{2}{c}{\textbf{--- Marginalised ---}}				& \textbf{5.07}	  &  \textbf{2.87} 	\\
	\hline \hline
	\end{tabular}
	\begin{tabular}{lccc}
	\multirow{2}{*}{Disk?} 	&	MW mass													&		\multicolumn{2}{l}{$m_\mathrm{WDM}/\mathrm{keV}>$}	\\
							&	$m = M_\mathrm{vir}^\mathrm{MW}/10^{12}M_\odot$			& 		1$\sigma$				& 2$\sigma$ 				\\
	\hline
	Y     					&   $0.8<m<1.0$ 											&		5.48     				&  3.07	 					\\ 
	Y     					&   $1.0<m<1.2$ 											&		4.80     				&  2.85	 					\\ 
	Y     					&   $1.2<m<1.7$ 											&		4.84     				&  2.69	 					\\ 
	\end{tabular}
\end{table}

Table~\ref{tab:WDM} shows how our constraints vary when we change various different model assumptions.
They are largely insensitive to the choice of WDM suppression index and choice of SMHM model (we do not investigate models with reduced halo occupation, for which the constraints would become stronger as satellites are necessarily hosted by less massive subhalos).
The lower section of Table~\ref{tab:WDM} shows how the results change when we perform the analysis splitting the simulated MW halos listed in Table~\ref{tab:halo_properties} into three mass bins.
If the MW has a mass below $10^{12}M_\odot$ \citep[favoured by e.g.][]{deason12,gibbons14} our constraints for the disk simulation strengthen to $m_\mathrm{WDM}>5.5(3.1)$ keV at 1$\sigma$(2$\sigma$).

\subsection{The prior on $m_\mathrm{WDM}$}
\label{sssec:wdm_prior}

Constraints on $m_\mathrm{WDM}$ derived from the Lyman-$\alpha$ forest have in the past been presented as likelihood distributions in $1/m_\mathrm{WDM}$ \citep{viel08,viel13,baur16}.
Interpreting these likelihoods distributions as posterior probability distributions, however, implicitly assumes a prior on the particle mass uniform in $1/m_\mathrm{WDM}$.
Such a prior is problematic since it is not invariant to rescaling of the problem and could potentially bias our inference.
A preferable prior would be $P(m_\mathrm{WDM})\propto1/m_\mathrm{WDM}$, which is invariant to rescalings.
This prior assigns equal probability to logarithmically spaced bins in $m_\mathrm{WDM}$, correctly describing our agnosticism on whether $m_\mathrm{WDM}/\mathrm{keV}$ is in the range 0.1-1, or 1-10, or 10-100, etc.

Having said that, when trying to constrain $m_\mathrm{WDM}$ with any probe of structure formation with some characteristic size, we are bound to lose sensitivity to increases in $m_\mathrm{WDM}$ beyond particle masses which suppress power on the scales of our probe.
In other words, constraints from structure formation can only ever provide a lower bound on $m_\mathrm{WDM}$.
The use of the uninformative prior $P(m_\mathrm{WDM})\propto1/m_\mathrm{WDM}$ would therefore lead to an improper posterior pdf on $m_\mathrm{WDM}$, with a divergent probability for high particle masses.
One way to circumvent this issue would be to restrict our attention to some finite range $m_\mathrm{WDM} < m_\mathrm{WDM}^\mathrm{max}$.
For some fixed choice of $m_\mathrm{WDM}^\mathrm{max}$ -- above which the system of interest looks sufficiently like CDM, say -- we could then report the constraints on the particle mass using the uninformative prior probability.
There remains some ambiguity in choosing $m_\mathrm{WDM}^\mathrm{max}$ however once some choice is made we can proceed with a correct statistical treatment rather than implicitly assume an undesired prior.
Taking $m_\mathrm{WDM}^\mathrm{max}=10$ keV, our marginalised constraints from Table~\ref{tab:WDM} become $m_\mathrm{WDM}>4.58(2.88)$ keV for the model without a disk, and  $m_\mathrm{WDM}>4.81(3.05)$ keV for the model with, both at 1$\sigma$(2$\sigma$) credibility.

\section{Discussion}
\label{sec:discuss}

\subsection{A robust test of halo occupation?}
\label{ssec:robust_test}

Our results show that the MW satellite galaxy luminosity function disfavors models where fewer than 10\% of $M_\mathrm{vir}=10^8 M_\odot$ halos host a luminous galaxy (see Section~\ref{subsubsec:ho_pplf}), which could be an important for constraint for models of reionisation.
This conclusion is robust against all SMHM variations we attempt -- i.e. using the fiducial, scatter or broken power-law -- and allowing the probability of halo occupation to fall as a power law with halo mass or allowing it to remain constant at 10\% below $M_\mathrm{vir}=10^8 M_\odot$ (see Figure~\ref{fig:halo_occupancy_compare}).
We now discuss whether this result may be sensitive to other aspects of our methodology or data-set.

Could our simulations simply not contain the population of subhalos required to host our sample of satellite galaxies?
We can imagine three reasons this may occur.
Firstly, we could be missing subhalos due to the limited resolution of our $N$-body simulations.
We check this by testing convergence of the halo mass functions with our two high-resolution runs -- one with a disk and one without -- which have particle masses 10 times smaller than our main simulation suite.
We find that at $z=0$ our main suite is well resolved above $M_\mathrm{vir}=3\times10^{7}M_\odot$ (corresponding to 132 particles), and hence if resolution is the reason we have under-predicted faint satellites, we must appeal to hosts less massive than this value to accommodate them.
In Appendix~\ref{sec:res} we show results running our entire inference machinery on the high-resolution simulations, finding that even these are unable to re-populate the faintest bin of the luminosity function.
This suggests that halos below $M_\mathrm{vir}=3\times10^{6}M_\odot$ would be required as hosts.

Secondly, we consider whether using a different halo finder may affect our results.
The comprehensive comparison project presented in \citet{onions12} found that a subhalo's mass can be reliably recovered - independent of halo finder - if it contains more than 100 particles.
Since this is fewer than the limit of 132 particles needed for mass function convergence, we believe that any changes resulting from use of a different halo finder will be smaller than changes when we go from using our medium- to our high-resolution simulations.
As shown in Appendix~\ref{sec:res}, such changes are small, and hence we believe that the majority of our results are robust against changing the halo finder.
Taking a slightly more detailed look, both \citet{onions12} and \citet{knebe13} both reveal that there is a spatial dependence to halo finder performance.
Both show that in the central regions ($r\lesssim40$ kpc) of the host, our chosen halo finder, \textsc{rockstar}, identifies more substructure than other commonly used alternatives, attributing this to the fact that only \textsc{rockstar} utilises phase space information.
For example, \textsc{subfind}, another popular choice, finds $\sim25\%$ less substructure in the host's centre.
Though there is no suggestion that the additional substructure found by \textsc{rockstar} is spurious, if this were the case, and instead we had used \textsc{subfind}, it would only act to strengthen our conclusion that halo occupation must be $>10\%$ at $10^8 M_\odot$.

Thirdly, some of the faint galaxies could have been accreted onto the MW as satellites of the LMC.
This is problematic, since the LMC is a rare occurrence in $\Lambda$CDM \citep{boylankolchin11}, thanks to its $\sim10^{11}M_\odot$ dark halo \citep{penarrubia16} and recent arrival to the MW \citep{kallivayalil13}.
Since we have not explicitly demanded that our simulated MW analogues host LMC analogues, its subhalo entourage will not be accounted for in our modeling.
To circumvent this issue, we have restricted our sample of faint satellites to those discovered in the SDSS-DR9, whose footprint north Galactic hemisphere is far from the LMC.
Using a dynamical model of LMC satellites constrained against the distribution of satellite galaxies in the DES survey \citep{jethwa16}, we calculate that the probability of LMC association for any of our sample of faint satellites is $p<0.05$, hence it is unlikely that any of the faint satellites in our sample are LMC contaminants.

We next consider two deficiencies in our model of satellite detection efficiency.
Firstly, we have used the binary selection function taken from \citet{koposov09}, i.e. our Equation~(\ref{eqn:selc_func}), however this is an approximation to a more complicated form calculated in \citet{koposov08}, in which detectability is not a binary choice but there is a spread in detection efficiency which becomes considerable (> 1 mag) at distances $D_\odot<30$ kpc (see figure 12 of \citealt{koposov08}, and also \citealt{walsh09}).
Accounting for this spread could feasibly inflate the uncertainty in our predicted luminosity functions.
Having said this, we note that the absolute number of subhalos with $r<30$ kpc is small -- in the range 1-5 (3-13) for our simulated halos with(out) a disk -- and at larger distances the assumption of a binary-threshold becomes valid.
Thus, any correction due to this effect is likely to be small.

Secondly, we have ignored the dependence of detection efficiency on surface brightness.
Instead, we have assumed that every galaxy brighter than the limiting magnitude also exceeds the 30 mag arcsec$^{−2}$ surface brightness SDSS detection threshold, since our model does not predict the size of the stellar system, only its luminosity.
In \citet{bullock10}, this size \textit{is} predicted using an empirically derived luminosity/velocity-dispersion relation for MW satellites, then applying the Jeans equation under the assumption that subhalos follow NFW profiles \citep{navarro97}.
Applying this to the Via Lactia II simulations \citep{diemand08}, they predict that there is a significant population of low-luminosity MW satellites with un-detectably low surface brightness.
Interestingly, subsequent, deeper surveys have consistently found many satellites around the 30 mag arcsec$^{−2}$ SDSS limit, with very few examples dropping below it \citep[e.g. the ultra-diffuse Crater 2,][]{torrealba16}.
Future discoveries of low-luminosity, low-surface brightness galaxies will only increase the requirement for high halo occupation fractions for low-mass halos.

A final possibility we consider is that the excess of observed faint satellites could be the result of halo-to-halo scatter in subhalo abundances.
We account for this in two ways: simulating six different MW analogues which inherently sample the scatter in subhalo abundances, and using a likelihood function which allows for Poisson scatter in the predicted number of galaxies per luminosity bin (see Section~\ref{ssec:probmod}).
\citet{boylankolchin10} quantify scatter in the number of subhalos contained within $R_\mathrm{vir}$ for the Millenium simulations \citep{springel05,boylankolchin09}, finding that for massive subhalos ($M_\mathrm{sub}/M_\mathrm{host}>3\times10^{-3}$) the scatter is indeed Poissonian about the mean, but at lower masses it becomes super-Poissonian, such that $\sigma=1.2\sigma_\mathrm{Poisson}$ for subhalos with masses $M_\mathrm{sub}/M_\mathrm{host}=10^{-3}$; thus, our inclusion of Poisson scatter may slightly underestimate the true scatter in subhalo abundance for low-mass halos.
We chose not to incorporate the \citet{boylankolchin10} result in our analysis, however, since it is valid for subhalo abundance within a virial radius and hence it may not be appropriate to assume super-Poissonian scatter for a local (i.e. observable) sub-sample of subhalos.
It is also unclear whether super-Poissonian scatter will apply once the destructive influence of the disk is taken into account, however such questions are beyond the scope of our current suite of simulations.

Lastly, we note that this result is statistically significant (see Bayes factors in Table~\ref{tab:bayes_factors}), however does rely on the observation of a handful of faint satellite galaxies above predictions.
As shown in Section~\ref{ssec:predit_survey} however, upcoming deep, large imaging surveys should be decisive in determining what fraction of low-mass halos host a galaxy, and the implications of this on the underlying model of reionisation.

\subsection{Implications for reionisation}

Our result that more than 10\% of MW subhalos with peak masses $M_\mathrm{vir}=10^8 M_\odot$ likely host a luminous galaxies lies in tension with two recent sets of hydrodynamic, cosmological, zoom-in simulations.
Of a suite of seven simulated field halos, \citet{shen2014} found that reionisation prevents any stars from forming in the three with $M_\mathrm{vir}<10^9M_\odot$.
\citet{sawala16} improved these statistics by simulating 12 Local Group analogues, each containing thousands of low-mass subhalos: they find that the fraction of $10^8 M_\odot$ halos which host a galaxy is $<10\%$ at all times and $\sim1\%$ at $z=0$.
Contast these results with two other recent works.
\citet{wheeler2015} form galaxies in halos with $z=0$ masses as low as $5\times10^8M_\odot$, though below $5\times10^9M_\odot$ these galaxies are uniformly composed of ancient, pre-reionisation stars.
The idealised simulations of \citet{blandhawthorn15} push the feasible mass of a galaxy-hosting halo down to $10^7M_\odot$.

The four works are listed in reverse-order of effective mass resolution, with the lowest resolution simulations finding that reionisation more strongly inhibits star formation.
This may be a consequence of high resolution being a prerequisite for the formation of dense substructures, which can self-shield against a front of ionising photons.
On the other hand, \citet{sawala16} show that -- despite using relatively low-resolution simulations over larger volumes -- halo occupancy fraction is converged between their medium to high-resolution runs, attributing this to the power of reionisation to limit star formation in low-mass halos at scales above the resolution limit of their simulations.
In addition to possible resolution effects, there remains significant uncertainty surrounding some of the physical parameters of reionisation, including the spectrum and escape fraction of ionising photons \citep{haardt12}.
Taken together, there seems enough uncertainty in current models of reionisation, and calculations of how this affects galaxy formation in low-mass halos, to accommodate our inferred constraint on $10^8 M_\odot$ halos.

\subsection{The SMHM relation: propogating and extrapolating uncertainties}

\begin{figure*}
\includegraphics[width=\textwidth]{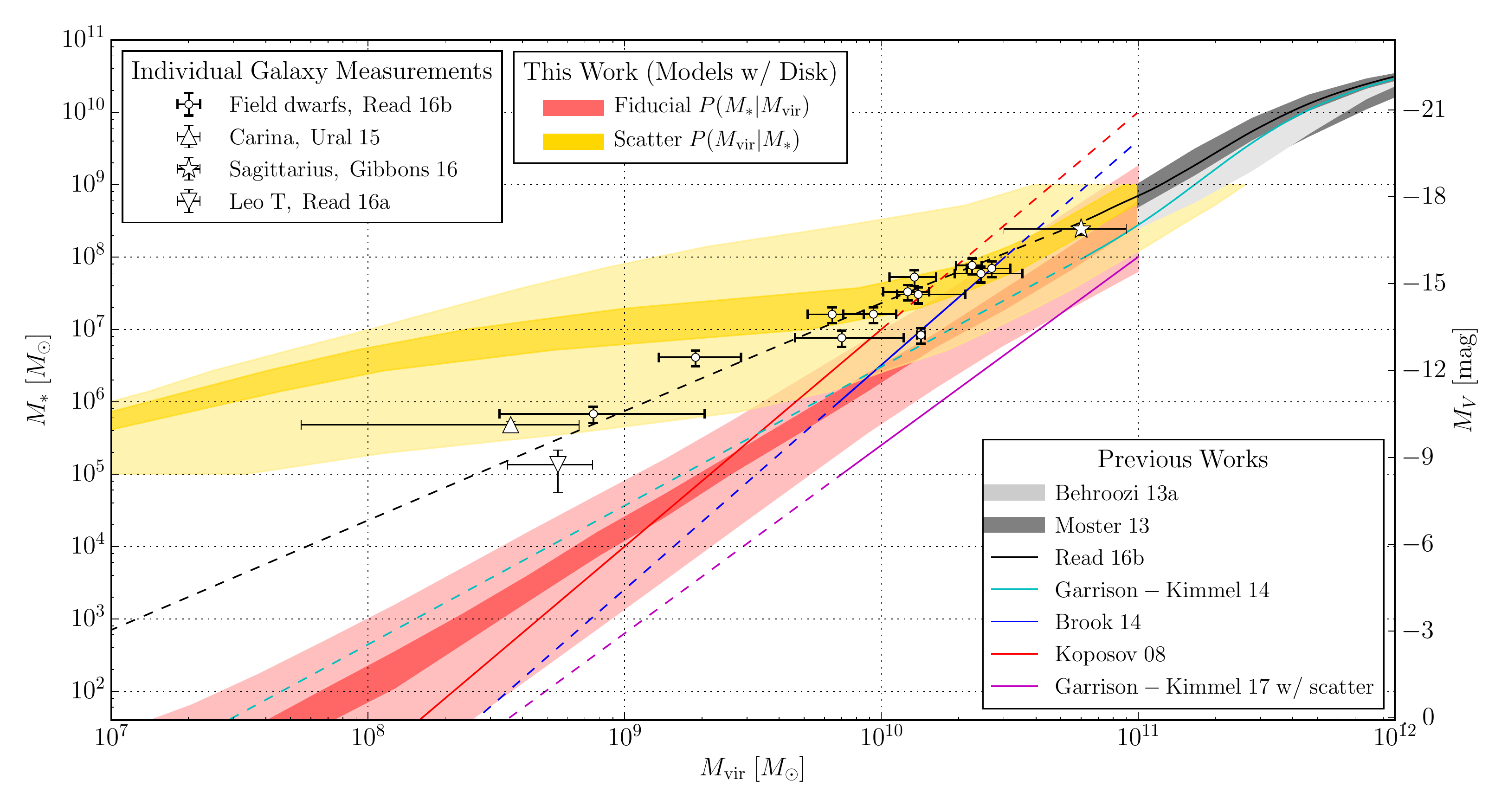}
\caption{
	Comparison of constraints/predictions of the SMHM relation.
	\textit{Previous works constraining the SMHM through abundance matching:}
	for halo masses $M_\mathrm{vir}>10^{11}M_\odot$ we show results from \citet{moster13}/\citet{behroozi13} using dark/light grey bands, the upper edges of which show the $z=0$ relations, the lower edges $z=3$.
	For $M_\mathrm{vir}<10^{11}M_\odot$ we show five previous results, using solid lines over the range where observational data was used and dashed lines for extrapolations: \citet{read16} in black,	\citet{garrison-kimmel_elvis} in cyan, \citet{brook14} in blue, \citet{garrison17} in magenta and \citet{koposov09} in red.
	\textit{Constraints from this works:}
	we show 68/95\% credibility intervals (dark/light coloured bands) on the posterior predictive SMHM relation for two of our present models, both of which include the effect of the MW disk.
	In red, we show constraints on stellar mass for a given halo mass (to be interpreted vertically) assuming our fiducial SMHM model; in yellow, constraints on halo mass for a given stellar mass (to be interpreted horizontally) assuming our scatter model.
	Allowing for scatter, constraints on halo mass at fixed stellar mass are significantly shallower than constraints on stellar mass at fixed halo mass.
	\textit{Individual measurements:}
	white symbols with errors show estimates for individual galaxies.
	For the Sagittarius \citep[star,][]{gibbons16} and Carina \citep[down triangle,][]{ural15} dwarf spheroidals, estimates come from $N$-body dynamical modelling.
	For Leo T (up triangle), from comparison to hydrodynamic simulations \citep{read16b}.
	For a sample of field galaxies \citep[cirles,][]{read16}, from fits to HI rotation curves.
	}
\label{fig:comparison}
\end{figure*}

Figure~\ref{fig:comparison} shows a comparison of inferences on the SMHM relation from abundance matching.
For $M_\mathrm{vir}>10^{11}M_\odot$ we show constraints from \citet{moster13} and \citet{behroozi13} (dark and light grey bands), which inform our prior on $P(M_*|M_\mathrm{vir}=10^{11}M_\odot)$ and hence match smoothly onto the constraints derived this work (red and yellow coloured bands).
For masses $M_\mathrm{vir}<10^{11}M_\odot$, the coloured lines show the results of five other previous works, shown as solid lines over the range where observational data-sets are used to constraint the models and dashed where they are extrapolations.
Three of these (\citealt{garrison-kimmel_elvis}, cyan; \citealt{brook14}, blue; \citealt{koposov09}, red) predict SMHM relations consistent with our constraints on stellar mass at fixed halo mass for our fiducial SMHM relation ($1\sigma$/2$\sigma$ constraints shown by dark/light red bands).
Two other works are discrepant: \citet{read16} predict a significantly shallower relation (black), while \citet{garrison17} a somewhat steeper one (magenta).
To understand why these differences arise, we briefly discuss the methodologies and data-sets used in these works.

Reassuringly, the only previous work to model the MW satellite galaxy luminosity function accounting for volume-incompleteness \citep[][red line]{koposov09} finds results similar to ours, despite their model being based on the extended Press-Schechter formalism rather than more realistic $N$-body simulations.
\citet{brook14} abundance match $M_*>10^6 M_\odot$ galaxies to constrained simulations of the Local Group and also find a steep SMHM relation (blue line), whereas the \citet{garrison-kimmel_elvis} relation (cyan line) is not directly found via abundance matching, but rather by modifying the \citet{behroozi13} relation to have a steeper faint-end slope.
This change is justified by propagating through a purpotedly improved observational constraint: replacing the SDSS derived \citet{baldry08} galaxy stellar mass function (complete above $M_*=10^{8.5}M_\odot$) for the \citet{baldry12} measurement from the Galaxy and Mass Assembly survey (GAMA) survey (complete above $M_*=10^8M_\odot$).
The result when making this change is that, when extrapolated to lower masses, the modified relation is consistent with Local Group observations, whereas the extrapolated \citet{behroozi13} relation significantly overpredicts the number of Local Group galaxies.

\begin{figure}
\includegraphics[width=\columnwidth]{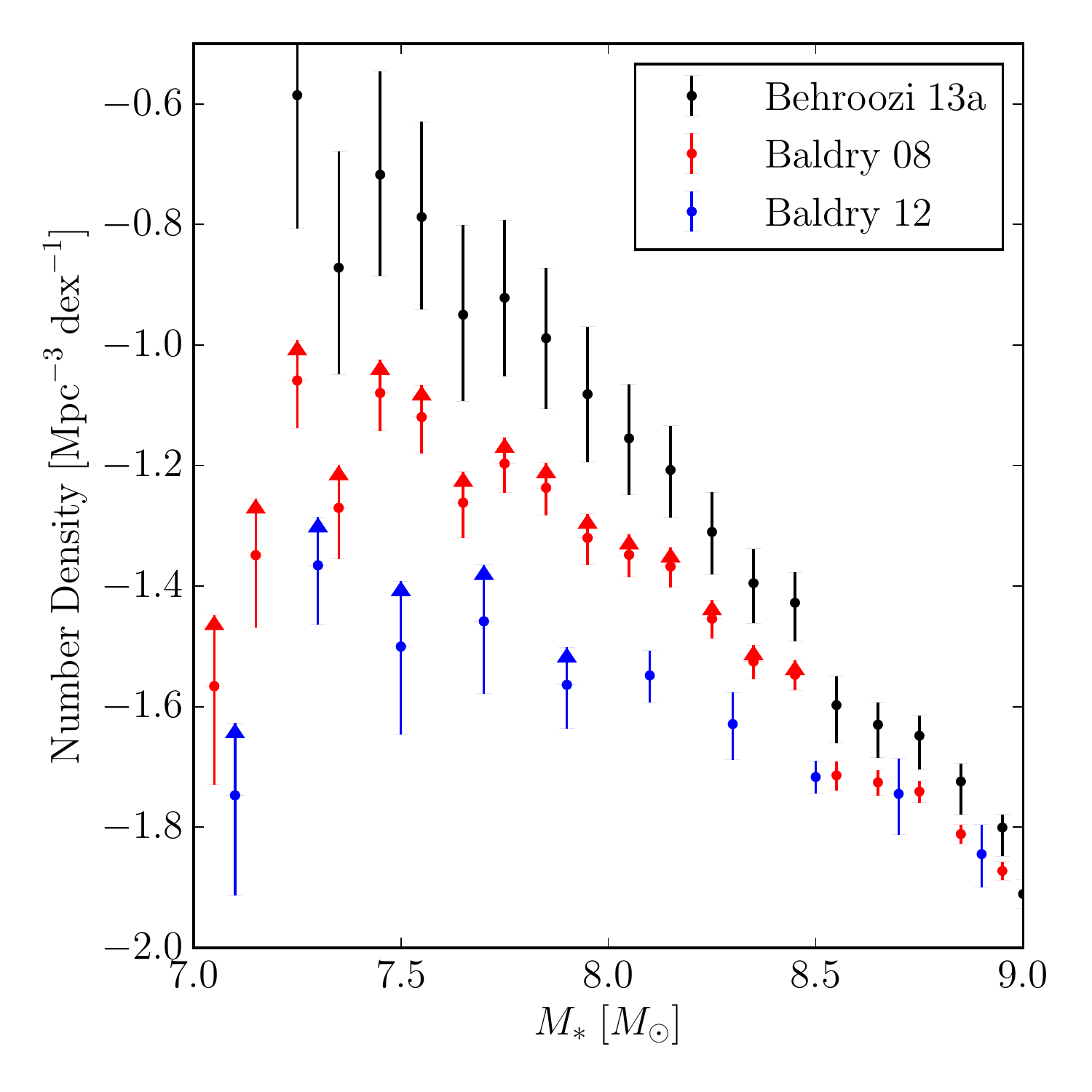}
\caption{
Galaxy stellar mass functions from \citet{baldry08} (red) and \citet{baldry12} (blue) where error bars are Poisson and up-arrows indicate lower-bounds due to surface brightness incompleteness.
Black points and error bars show the \citet{behroozi13} stellar mass function which applies a surface brightness correction to the \citet{baldry08} data.
}
\label{fig:GSMFs}
\end{figure}

For their abundance matching study, \citet{read16} use the same observational data-set as \citet{behroozi13}, finding a shallow inferred SMHM relation (black line).
The discussion in \citet{garrison-kimmel_elvis} suggests that if \citet{read16} were to use the \citet{baldry12} rather than the \citet{baldry08} galaxy stellar mass function, this relation would likely steepen.
\citet{read16} concede this point, however counter that since the survey volume of the GAMA survey is only one tenth that of SDSS, the \citet{baldry08} stellar mass function is more prone to cosmic variance and hence not necessarily an improvement on using \citet{baldry12}.
\citet{read16} also argue that a shallow SMHM relation may be made consistent with the Local Group galaxy abundance if environmental factors re-shape the field-derived SMHM relation, alleviating the tensions found by \citet{garrison-kimmel_elvis}.

To complicate matters further, we note that \citet{moster13} use the \citet{baldry08} mass function but their SMHM relation has a \textit{steep} faint-end slope much like \citet{garrison-kimmel_elvis}.
Why has the same data-set given differing results?
In Figure~\ref{fig:GSMFs} we compare the faint-ends of the galaxy stellar mass functions from \citet[][red symbols]{baldry08} and \citet[][blue symbols]{baldry12}, showing that \citet{baldry08} is indeed steeper.
The stellar mass function used by \citet{behroozi13} and \citet{read16}, however, applies a correction for incompleteness due to surface-brightness (black symbols).
Propagating uncertainties in this correction into the abundance matching fit would lead to a very uncertain prediction of the SMHM relation when extrapolated to low masses \citep[this may also be true for corrections due to large scale structure, e.g.][]{efstathiou88}.
The lack of any incompleteness correction may explain way \citet{moster13} predicts a steep faint-end SMHM slope.
Another explanation is that the relative inflexibility of the 4-parameter \citet{moster13} SMHM model makes it less sensitive to the faint-end slope than the more complex models of \citet{behroozi13} or \citet{read16}.

The matters of both propagating uncertainties into the analysis and answering questions of how much complexity is warranted by the data are best answered in a probabilistic, Bayesian framework, e.g. as we describe in Section~\ref{ssec:probmod}.
We believe that the apparent discrepancy between \citet{read16} and other works would become less significant if uncertainties in the stellar mass function were propagated into the analysis.
This is especially important if we extrapolate a SMHM relation orders of magnitude below the range of the observational data used to constrain it.

\subsection{The SMHM relation: comparing at fixed stellar mass or halo mass?}

We now consider measurements for individual galaxies (Figure~\ref{fig:comparison}, white symbols with error bars), which lie systematically above our constraints on on stellar mass at fixed halo mass (red coloured bands).
The halo mass estimates for these galaxies come from various sources.
The majority are inferred from HI rotation curves fit using a mass model for the dark matter halo which accounts for cusp-core transformations due to stellar feedback \citep[white circles, also from][where we have corrected the small difference between their $M_{200}$ compared to our use of $M_\mathrm{vir}$.]{read16}.
Two estimates -- one the low-mass end \citep[Carina, up-triangle,][]{ural15} and one at the high-mass end \citep[Sagittarius, star symbol,][]{gibbons16} -- come from $N$-body modelling of tidal features.
The estimate for the least luminous galaxy comes from direct comparison to the properties of dwarf galaxies produced in hydrodynamic simulations \citep[Leo T, down-traingle,][]{read16b}.
Considered altogether, these measurements suggest a shallow SMHM relation with little scatter, seemingly inconsistent with our inferred fiducial constraints on $P(M_*|M_\mathrm{vir})$.

We believe that scatter in the SMHM relation could be behind this seeming discrepancy.
This conclusion would appear at odds with the results of \citet{garrison17} (magenta line) who find that the inclusion of scatter leads to a steepening of the inferred stellar mass for a fixed halo mass.
Similarly, we also find that constraints on $P(M_*|M_\mathrm{vir})$ are steepest for the model with scatter (see Section~\ref{subsubsec:scatter_smhm}).
However, since galaxies are observed subject to their stellar mass, we should not expect them to sample the inferred $P(M_*|M_\mathrm{vir})$ distribution, but rather that of halo mass for a given stellar mass, i.e. $P(M_\mathrm{vir}|M_*)$.
Looking at such constraints for the scatter model (Figure~\ref{fig:comparison}, yellow coloured bands) we see good agreement between our inferred distribution and the individual galaxy measurements for stellar masses $M_*>10^{7.5}M_\odot$.
This agreement breaks down for fainter galaxies where the amount of scatter permitted by our model has diluted the relation between halo mass and stellar mass to the extent that we simply infer that the most abundant halos (i.e. the least massive) are the most likely hosts.
Our model is likely broken in this regime, however we can surmise an important point: scatter in the SMHM relation can reconcile small halo masses measured for individual galaxies with steep SMHM relations inferred from abundance matching, provided that we differentiate between $P(M_*|M_\mathrm{vir})$ and $P(M_\mathrm{vir}|M_*)$.
Our current implementation of SMHM scatter -- symmetric, log normal, growing linearly with $\log M_\mathrm{vir}^{-1}$ -- is somewhat ah-hoc, however in future works we plan to investigate more prescriptive, physically motivated models which may help to tighten the low-mass constraints.

\subsection{Warm Dark Matter constraints: comparison to previous works}

Even before accounting for disk destruction, our constraint on the mass of the warm dark matter (WDM) particle -- $m_\mathrm{WDM}>4.6(2.7)$ keV at 1$\sigma$(2$\sigma$) -- is stronger than other previous constraints placed with MW satellites: $m_\mathrm{WDM}>1$ keV from \citet{maccio10dm}, $m_\mathrm{WDM}>2.3$keV (2$\sigma$) from \citet{polisensky11}, and $m_\mathrm{WDM}>1.6$ keV from \citet{lovell14}.
We are able to place stronger constraints than these works since we account for the fact that the faintest satellites are only observed locally.
Once we include the effect of the baryonic disk, our lower-bound on the mass of the warm dark matter particle -- $m_\mathrm{WDM}>5.1(2.9)$ keV at 1$\sigma$(2$\sigma$) -- becomes competitive with the most recent constraint of $m_\mathrm{WDM}>3.0$ keV at 2$\sigma$ coming from the Lyman-$\alpha$ forest \citep[][rising to $m_\mathrm{WDM}>4.1$ keV when they omit the Planck 2016 prior on the spectral index of primordial density fluctuations]{baur16}. We therefore deduce that, pre-infall, the faintest MW satellite was hosted by a dark matter halo at most as massive as the structures probed by the narrowest Lyman-$\alpha$ absorption features.

\citet{kennedy14} investigate the effect of changing the assumed MW mass on the inferred WDM constraints, finding $m_\mathrm{WDM}>20$ keV for $M_\mathrm{MW}<10^{12}M_\odot$, altogether ruling out some WDM models \citep[e.g.][]{perez16}.
This constraint, however, is extremely sensitive to the parameters of their galaxy formation model, in particular the minimum $v_\mathrm{max}$ for a subhalo to be able to cool gas and form a galaxy: changing this from 30 km/s to 25 km/s, their lower bound on $m_\mathrm{WDM}$ decreases by a factor of 10.
Note that both of these choices of $v_\mathrm{max}$ correspond to peak halo masses $M_\mathrm{vir}>10^{9}M_\odot$, and hence using these as a threshold for galaxy formation will underpredict the abundance of observed MW ultra-faints, as shown in Section~\ref{subsubsec:ho_pplf}.
Our WDM constraints fit the observed luminosity function, simultaneously fitting for the WDM particle mass and parameters of the SMHM relation, rather than depending on some prescribed model of galaxy formation.
Splitting our results by MW mass (see Table~\ref{tab:WDM}), we find that for $M_\mathrm{MW}<10^{12}M_\odot$ \citep[as favoured by e.g.][]{gibbons14,deason12}, our constraints improve to $m_\mathrm{WDM}>5.5(3.1)$ keV.

One possible criticism is that our WDM constraints are derived from \textit{post processed} CDM simulations, where we include the depletion in subhalo abundance but ignore other effects of changing from cold to warm dark matter.
These include, for example, the fact that WDM halos have central density cores compared to CDM cusps \citep{tremaine79} with cored subhalos being tidally destroyed more easily than cusped CDM halos \citep{errani_etal_2017}.
For the allowed range of WDM particle masses, however, cores are created on parsec scales \citep{maccio12,shao13} and hence not dynamically relevant.
Independent of the central density slope, \citet{bose16} show that subhalo central densities can be reduced by 10-30\% in a WDM versus CDM universe, which would lead to more tidal disruption of subhalos; accounting for this would strengthen our constraints.
We finally note that \citet{lovell14} show that the radial distribution of subhalos is unchanged between $\Lambda$CDM and $\Lambda$WDM.

\section{Conclusions}
\label{sec:conclude}

In this work, we have investigated the connection between galaxies and dark matter halos at the lowest-mass scales.
We do this by modelling the luminosity function of MW satellite galaxies using parameterised Stellar-Mass-Halo-Mass (SMHM) relations of varying complexity.
We apply these to two suites of cosmological, zoom-in simulations of MW like halos: one $N$-body only, and one with an analytic disc grown in the host halo.
From this we build a probabilistic model, which we constrain using against the observed MW satellite luminosity function.
In contrast to other similar works, by taking into account the observational biases we push the model/data comparison to the faintest known galaxies, which allows us to draw the following conclusions:

\begin{enumerate}
\item Assuming all subhalos host a galaxy, the luminosity function of MW satellite galaxies can be well modelled by a single power-law SMHM relation with or without scatter (top row of Figure~\ref{fig:pplf_disk}) with a slope consistent with the steeper-end of previous estimates (red coloured bands in Figure~\ref{fig:comparison}).
\item Restricting the fraction of subhalos with peak masses of $10^8M_\odot$ to less than 10\%, we underpredict the number of very faint ($M_V>-4$) MW satellites, even if we allow for significant scatter in the SMHM relation, or a characteristic scale below which the SMHM relation becomes shallower (bottom row of Figure~\ref{fig:pplf_disk}).
\item Our preference for halo occupation fractions greater than 0.1 is a statistically positive but small (i.e. Bayes factor $>3$, Table~\ref{tab:bayes_factors}) however future surveys should provide a decisive answer (see Figure~\ref{fig:predict_surveys} for predictions).
If the ongoing Subaru/Hyper Suprime-Cam survey discovers more than seven $M_V<-1$ MW satellite galaxies, or more than two in the faintest known regime $-1<M_V<0$, this would be further positive evidence for high occupation fractions.
\item Including scatter in the SMHM relation steepens the inferred relation of stellar mass for a fixed halo mass (due to an Eddington bias, see Section~\ref{sssec:scatter}), but it gives a shallower relation between inferred halo mass for a fixed stellar mass, assuming a $\Lambda$CDM prior on the halo mass function.
This is because scatter allows low-mass halos to host bright galaxies and, even if this possibility is unlikely, because low-mass halos are more abundant they become the likely hosts over a wide range of stellar mass (see Section~\ref{ssec:invert_constraints}).
This may be an important consideration when comparing measurements for individual galaxies -- which are observed subject to their stellar mass - to results from abundance matching.
\item Our most conservative upper-bound for the pre-infall virial mass of Segue I -- i.e. the faintest galaxy in our sample -- is $2.4(13.9)\times10^8M_\odot$ at $1\sigma$(2$\sigma$).
Our upper bound on the $z=0$ mass are $0.6(2.1)\times10^8M_\odot$ at $1\sigma$(2$\sigma$) (see Figure~\ref{fig:mass_lightest}).
\item Though the MW disk does effectively destroy subhalos in the inner region of the halo, we find that using slightly shallower SMHM relation (i.e. giving a Segue I halo mass $\sim0.2$ dex lighter for a disk model compared to a no-disk model) compensates for this effect when trying to reproduce the observed satellite luminosity function.
\item Translating our experiment to a Warm Dark Matter (WDM) cosmology, we constrain the mass of the WDM particle to $m_\mathrm{WDM}>5.1(2.9)$ keV at 1$\sigma$(2$\sigma$) (see Section~\ref{sec:wdm}).
Imposing a prior that the virial mass of the MW is less than $10^{12}M_\odot$, these strengthen to $m_\mathrm{WDM}>5.5(3.1)$ keV.
\item All of the numerical constraints listed here are achieved when we include the MW disk and its ability to destroy Galactic subhalos.
Throughout this work, we quantify the effect of ignoring the disk destruction, e.g. it leads to weaker WDM constraints $m_\mathrm{WDM}>4.6(2.7)$ keV.
\end{enumerate}

\section*{acknowledgements}

We thank Sergey Koposov for writing the tools to ingest the simulations, halo catalogs, and merger trees into an SQL database.
We also thank Debora Sijacki and Ewald Puchwein for helpful advice for using and modifying \textsc{gadget-3}.
PJ thanks the Science and Technology Facilities Council for the award of a studentship, and Justin Read, Shea Garrison-Kimmel, Alis Deason, Marius Cautun, Sownak Bose and Carlos Frenk for helpful discussions, and the anonymous referee for helpful comments.
The research leading to these results has received funding from the European Research Council under the European Union's Seventh Framework Programme (FP/2007-2013) / ERC Grant Agreement n. 308024.





\footnotesize{ \bibliographystyle{mn2e} \bibliography{mybib} }




\appendix

\section{Resolution Tests}
\label{sec:res}

\begin{figure*}
\includegraphics[width=\textwidth]{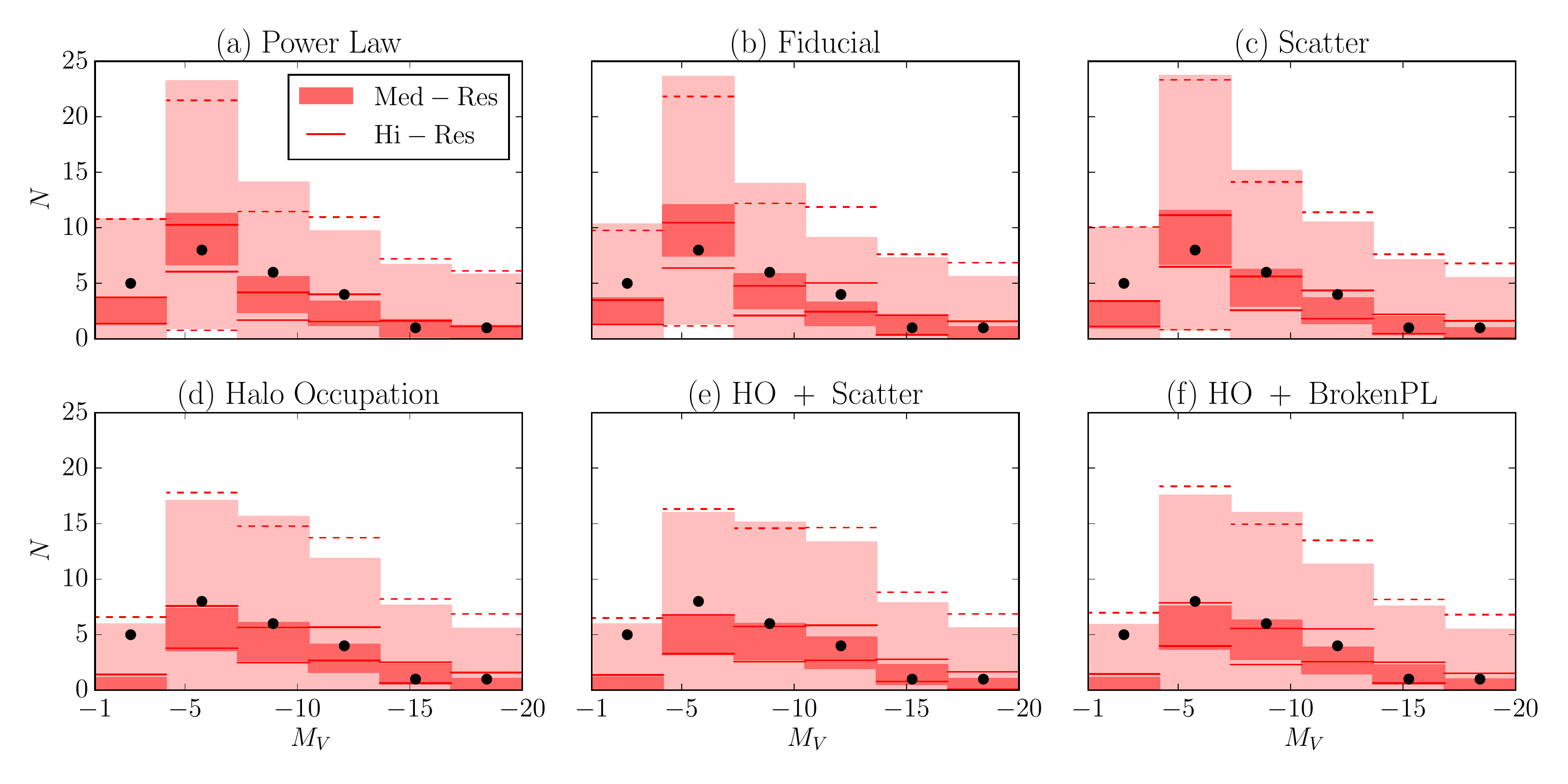}
\caption{
Resolution test for luminosity function, for simulations with a disk.
The luminosity functions from the medium resolution simulation are shown by the coloured bands, those from the equivalent high resolution simulation by the solid/dashed lines.
}
\label{fig:res_pplf}
\includegraphics[width=\textwidth]{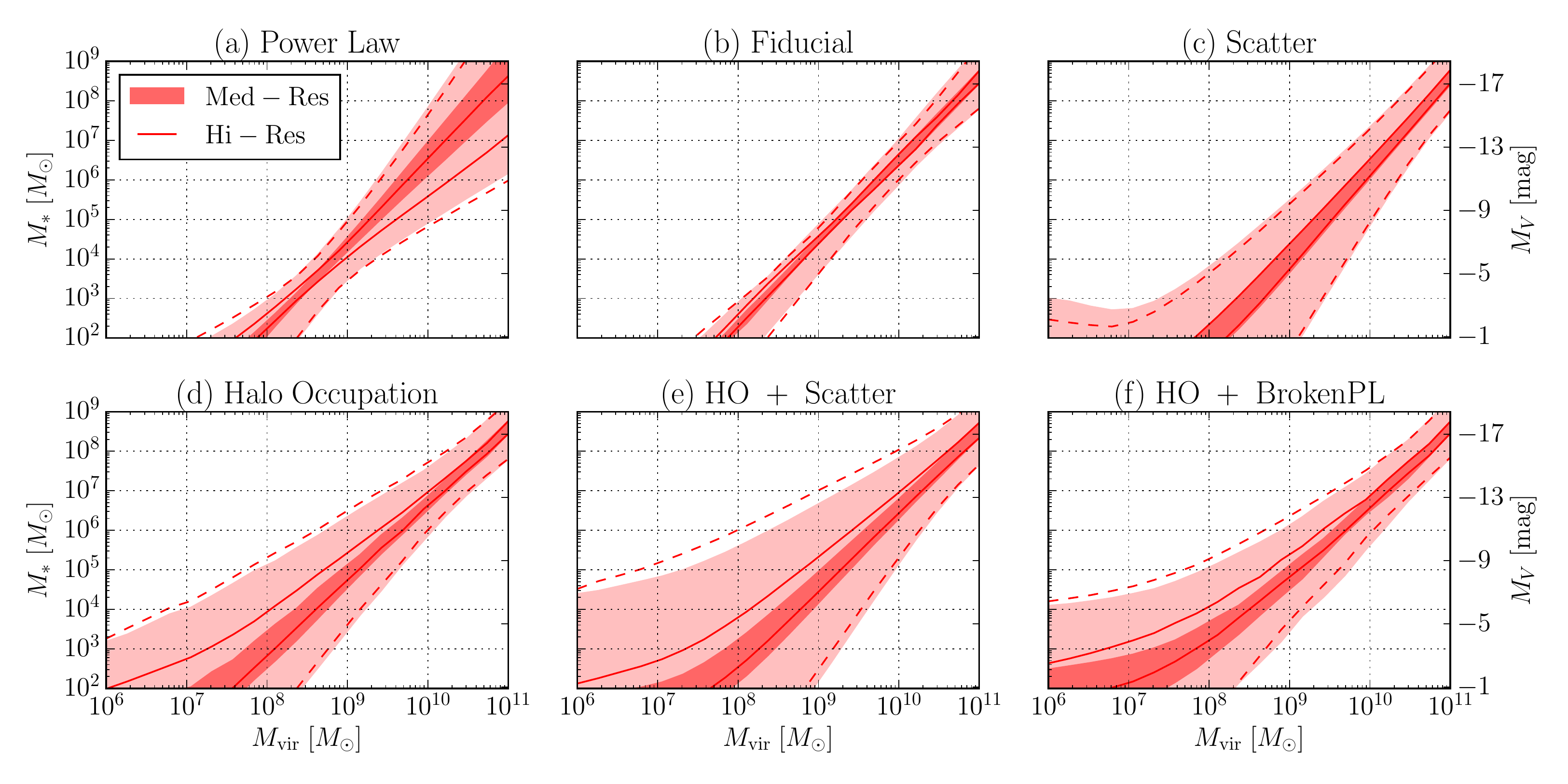}
\caption{
Resolution test for $P(M_*|M_\mathrm{vir},\mathbf{X})$, shown for simulations with a disk. Constraints from the medium resolution simulation are shown by the coloured band, while those from the equivalent high resolution simulation by the solid/dashed lines.
}
\label{fig:res_smhm}
\end{figure*}

For the analysis presented in the main body of this paper we use 7 halos simulated at medium-resolution (each once with a disk, once without) using a particle mass $M_p=1.54\times10^{5}h^{-1}M_\odot$
One of those halos has also been simulated at high-resolution, with a particle mass $M_p=1.92\times10^{4}h^{-1}M_\odot$.
We mainly do this to check convergence of the halo mass function, however, here we run our entire inference machinery on the high resolution simulations to see how the results are affected.
Figures~\ref{fig:res_pplf} show the posterior predictive luminosity function (see Section~\ref{ssec:posterior_predictive}), while \ref{fig:res_smhm} shows the posterior predictive distribution on stellar mass for a given halo mass (see Section~\ref{ssec:posterior_predictive}).
These are shown for the six models of SMHM relation and halo occupancy described in Section~\ref{ssec:mstarmods}, in both cases for simulations including the MW disk.

In both Figures, the coloured bands show that results from the medium-resolution simulation, while the solid/dashed lines show the results from the high-resolution version.
For all models shown, the results are converged between the two resolution levels.
Importantly, the high resolution simulation (which have halo mass functions converged to $M_\mathrm{vir}=3\times10^{6}M_\odot$) has not been able to reconcile the model/data tension in the faintest luminosity bin for models with restrivted halo occupation (bottom row of Figures~\ref{fig:res_pplf}).
This suggests that our results regarding the halo occupation fraction are not sensitive to the numerical resolution of our $N$-body simulations.


\bsp	
\label{lastpage}
\end{document}